\pdfoutput=1

\documentclass[sigconf,10pt]{acmart}

\usepackage{booktabs} 

\usepackage{url}
\usepackage{graphicx}
\usepackage{balance}  
\usepackage{xcolor}
\usepackage{subcaption}
\usepackage{caption}
\usepackage[ruled,vlined,linesnumbered]{algorithm2e}
\usepackage[noend]{algpseudocode}
\usepackage{paralist}
\usepackage{listings}
\usepackage{fancyvrb}
\usepackage{mathtools}
\usepackage{amsmath}
\usepackage{enumitem}
\usepackage[subtle,title=tight]{savetrees}

\renewcommand\footnotetextcopyrightpermission[1]{} 
\AtBeginDocument{%
  \providecommand\BibTeX{{%
    \normalfont B\kern-0.5em{\scshape i\kern-0.25em b}\kern-0.8em\TeX}}}

\newcommand\srm[1]{\textcolor{red}{}}
\newcommand\anil[1]{\textcolor{blue}{}}
\newcommand\yxy[1]{\textcolor{green}{}}

\newcommand\rev[1]{#1}
\newcommand\new[1]{#1}

\setcopyright{none}
\copyrightyear{2020}
\acmYear{2020}
\acmDOI{XXX-XXX-XXX}

\copyrightyear{2020}
\acmYear{2020}
\setcopyright{acmcopyright}
\acmConference[SIGMOD '20] {2020 ACM SIGMOD International Conference on Management of Data}{June 14--19, 2020}{Portland, OR, USA}
\acmBooktitle{2020 ACM SIGMOD International Conference on Management of Data (SIGMOD'20), June 14--19, 2020, Portland, OR, USA}
\acmPrice{15.00}
\acmDOI{10.1145/3318464.3380595}
\acmISBN{978-1-4503-6735-6/20/06}

\begin{document}
\fancyhead{}

\sloppy

\title{A Study of the Fundamental Performance Characteristics of GPUs and CPUs for Database Analytics \\ (Extended Version)}



\author{Anil Shanbhag}
\affiliation{\institution{MIT}}
\email{anil@csail.mit.edu}

\author{Samuel Madden}
\affiliation{\institution{MIT}}
\email{madden@csail.mit.edu}

\author{Xiangyao Yu}
\affiliation{\institution{University of Wisconsin-Madison}}
\email{yxy@cs.wisc.edu}

\renewcommand{\shortauthors}{}

\begin{abstract}
There has been significant amount of excitement and recent work on GPU-based
database systems. Previous work has claimed that these systems can perform
orders of magnitude better than CPU-based database systems on analytical
workloads such as those found in decision support and business intelligence
applications. A hardware expert would view these claims with suspicion. 
Given the general notion that database operators are memory-bandwidth bound, one
would expect the maximum gain to be roughly equal to the
ratio of the memory bandwidth of GPU to that of CPU. 
In this paper, we adopt a model-based approach to understand when and why the
performance gains of running queries on GPUs vs on CPUs vary from the
bandwidth ratio (which is roughly 16$\times$ on modern hardware). We propose Crystal, a library of parallel routines that can
be combined together to run full SQL queries on a GPU with minimal
materialization overhead. We implement individual query operators to show that
while the speedups for selection, projection, and sorts are near the bandwidth ratio, joins
 achieve less speedup due to differences in
hardware capabilities. Interestingly, we show on a
popular analytical workload that full query performance gain from running on
GPU exceeds the bandwidth ratio despite individual operators having speedup
less than bandwidth ratio, as a result of limitations of vectorizing chained operators
on CPUs, resulting in a 25$\times$ speedup for GPUs over CPUs on the benchmark.
\end{abstract}






\maketitle

\section{Introduction}

In the past decade, special-purpose graphics processing units (GPUs) have evolved
into general purpose computing devices, with the advent of general purpose
parallel programming models, such as CUDA~\cite{cuda} and OpenCL~\cite{opencl}. Because
of GPU's high compute power, they have seen significant adoption in
deep learning and in high performance computing~\cite{gpu-hpc}. GPUs also have significant
potential to accelerate memory-bound applications such as database systems. GPUs utilize High-Bandwidth Memory
(HBM), a new class of RAM that has significantly higher throughput compared
to traditional DDR RAM used with CPUs. A single modern GPU can have up to
32~GB of HBM capable of delivering up to 1.2 TBps of memory bandwidth and 14 Tflops
of compute. In contrast, a single CPU can have hundreds of GB of memory with up
to 100 GBps memory bandwidth and 1 TFlop of compute.

This rise in memory capacity, coupled with the ability to equip a modern server with several GPUs (up to 20), means that it's possible to have hundreds of gigabytes of GPU memory on a modern server.  This is sufficient for many analytical tasks;  for example, one machine could host several weeks of a large online retailer's (with say 100M sales per day) sales data (with 100 bytes of data per sale) in GPU memory, the on-time flight performance of all commercial airline flights in the last few decades, or the time, location, and (dictionary encoded) hash tags used in every of the several billion tweets sent over the past few days.


In-memory analytics is typically memory bandwidth bound. The improved memory bandwidth of GPUs has led some researchers to use GPUs as coprocessors for analytic query processing~\cite{yinyang,gpudb2,hippogriffdb,funke2018pipelined}. However, previous work leaves several  unanswered questions:
\vspace{-0.05in}
\begin{itemize} [leftmargin=*]
\item GPU-based database systems have reported a wide range of performance improvement compared to CPU-based database systems, ranging from $2\times$ to $100\times$. There is a lack of consensus on how much performance improvement can be obtained from using GPUs. Past work frequently compares  against inefficient baselines, e.g., MonetDB~\cite{yinyang,gpudb2,hippogriffdb}  which is known to be inefficient~\cite{hyper}. The empirical nature of past work makes it hard to generalize results across hardware platforms.

\item Past work generally views GPUs strictly as an coprocessor. Every query ends up shipping data from CPU to GPU over PCIe. Data transfer over PCIe is an order of magnitude slower than GPU memory bandwidth, and typically less than the CPU memory bandwidth.
As a result, the PCIe transfer time becomes the bottleneck and limits gains. To the extent that past work shows performance improvements using GPUs as an coprocessor, much of those gains may be  due  to evaluation against inefficient baselines.

\item There has been significant improvement in GPU hardware in recent years. Most recent work on GPU-based database~\cite{funke2018pipelined} evaluates on GPUs which have memory capacity and bandwidth of 4 GB and 150 GBps respectively, while latest generation of GPUs have $8\times$ higher capacity and bandwidth. These gains significantly improve the attractiveness of GPUs for query processing.
\end{itemize}
\vspace{-0.05in}



In this paper, we set out to understand the true nature of performance difference between CPUs and GPUs, by performing rigorous model-based and performance-based analysis of database analytics workloads after applying optimizations for both CPUs and GPUs. To ensure that our implementations are state-of-the-art, we use theoretical minimums derived assuming memory bandwidth is saturated as a baseline, and show that our implementations can typically saturate the memory bus, or when they cannot, describe in detail why they fall short.  Hence, although we offer some insights into the best implementations of different operators on CPUs and GPUs, the primary contribution of this paper is to serve as a guide to implementors as to what sorts of performance differences one should expect to observe in database implementations on modern versions of these different architectures.

Past work has used GPUs mainly as coprocessors. By comparing an efficient CPU implementation of a query processor versus an implementation that uses the GPU as a coprocessor, we  show that GPU-as-coprocessor offers little to no gain over a pure CPU implementation,  performing worse than the CPU version for some queries. We argue that the right setting is having the working set stored directly on GPU(s).

We developed models and implementations of \new{basic operators: Select, Project, and Join} on both CPU and GPU 
to understand when the ratio of operator runtime on CPUs to runtime on GPUs deviates from the ratio of memory bandwidth of GPU to memory bandwidth of CPU. In the process, we noticed that the large degree of parallelism of GPUs leads to additional materialization. We propose a novel execution model for query processing on GPUs called the \textit{Tile-based execution model}. Instead of looking at GPU threads in isolation, we treat a block of threads (``thread block'') as a single execution unit, with each thread block processing a tile of items. The benefit of this tile-based execution model is that thread blocks can now cache tiles in shared memory and collectively process them. This helps avoid additional materialization. This model can be expressed using a set of primitives where each primitive is a function which takes as input of set of tiles and outputs a set of tiles. We call these primitives \textit{block-wide functions}. We present \texttt{Crystal}, a library of block-wide functions that can be used to implement the common SQL operators as well as full SQL queries. Furthermore, we use \texttt{Crystal} to implement the query operators on the GPU and compare their performance against equivalent state-of-the-art implementations on the CPU.
We use \texttt{Crystal} to implement the Star-Schema Benchmark (SSB)~\cite{ssb} on the GPU and compare it's performance against our own CPU implementation, a state-of-the-art CPU-based OLAP DBMS and a state-of-the-art GPU-based OLAP DBMS. In both cases, we develop models assuming memory bandwidth is saturated and reason about the performance based on it.

In summary, we make the following contributions:
\vspace{-0.05in}
\begin{itemize}[leftmargin=*]
\item We show that previous designs which use the GPU as a coprocessor show no performance gain when compared against a state-of-the-art CPU baseline. Instead, using modern GPU's increased memory capacity to store working set directly on the GPU is a better design.
\item We present \texttt{Crystal}, a library of data processing primitives that can be composed together to generate efficient query code that can full advantage of GPU resources.
\item We present efficient implementations of individual operators for both GPU and CPU. For each operator, we provide cost models that can accurately predict their performance. 
\item We describe our implementation of SSB and evaluate both GPU and CPU implementations of it. We present cost models that can accurately predict query runtimes on the GPU and discuss why such models fall short on the CPU.
\end{itemize}
\vspace{-0.05in}

\section{Background}

In this section, we review the basics of the GPU architecture and describe relevant aspects of past approaches to running database analytics workloads on CPU and GPU.

\subsection{GPU Architecture}
\label{sec:gpu-memory-hierarchy}

Many database operations executed on the GPU are performance
bound by the memory subsystem (either shared or global
memory)~\cite{yinyang}. In order to characterize the performance of
different algorithms on the GPU, it is, thus, critical to properly
understand its memory hierarchy.

\begin{figure}[t]
\centering
\includegraphics[width=0.9\columnwidth]{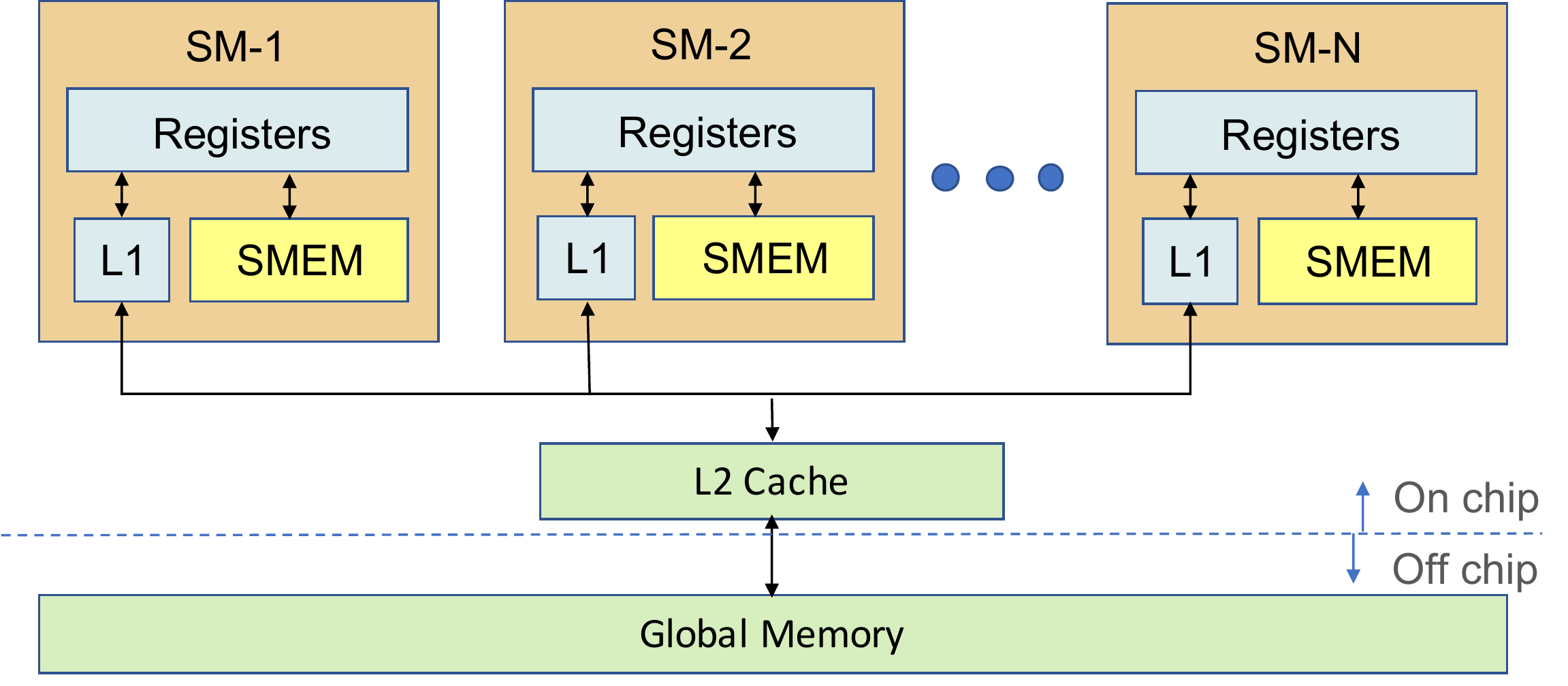}
\vspace{-.1in}
\caption{GPU Memory Hierarchy}
\label{fig:memory-hierarchy}
\vspace{-.15in}
\end{figure}

Figure~\ref{fig:memory-hierarchy} shows a simplified hierarchy of a modern GPU.
The lowest and largest memory in the hierarchy is the \textit{global memory}. 
A modern GPU can have global memory capacity of up to 32~GB with memory bandwidth of up to 1200 GBps.
Each GPU has a number of compute units called \textit{Streaming Multiprocessors (SMs)}.
Each SM has a number of cores and a fixed set of registers. Each SM also has a
\textit{shared memory} which serves as a scratchpad that is controlled by the programmer and can be accessed by all the cores in the SM. 
Accesses to global memory from a SM are cached in the L2 cache (L2 cache is shared across all SMs) and
optionally also in the L1 cache (L1 cache is local to each SM).

Processing on the GPU is done by a large number of threads organized into
{\it  thread blocks} (each run by one SM). Thread blocks are further divided
into groups of threads called warps (usually consisting of 32 threads). The threads of a warp execute in a
\textit{Single Instruction Multiple Threads (SIMT)} model, where each thread executes the same instruction stream on different data. The device
groups global memory loads and stores from threads in a single warp
such that multiple loads/stores to the same cache line are combined into a single request.
Maximum bandwidth can be achieved when a warp's access to
global memory results in neighboring locations being accessed.

The programming model allows users to explicitly allocate global memory and
shared memory in each thread block. Shared memory has an order of magnitude higher bandwidth
than global memory but has much smaller capacity (a few MB vs. multiple GB).

Finally, registers are the fastest layer of the memory hierarchy. If a
thread block needs more registers than available, register values
spill over to global memory.

\subsection{Query Execution on GPU}

With the slowing of Moore's Law, CPU performance has stagnated. In recent years,
researchers have started exploring heterogeneous computing to overcome the scaling problems
of CPUs and to continue to deliver interactive performance for database applications.
In such a hybrid CPU-GPU system, the two processors are connected via PCIe. 
The PCIe bandwidth of a modern machine is up to 16~GBps, which is much lower than the memory bandwidth of either CPU or GPU. Therefore, data transfer between CPU and GPU is a serious performance bottleneck. 


Past work in the database community has focused on using the GPU as a coprocessor, which we call the \textit{coprocessor model}. In this model, data primarily resides in CPU's main memory. For query execution, data is shipped from the CPU to the GPU over PCIe, so that (some) query processing can happen on the GPU. Results are then shipped back to the CPU. 
Researchers have worked on optimizing various database operations under the co-processor model: selection~\cite{gpu-select1}, join~\cite{gpu-join1,gpu-join2,gpu-join3,sitaridi2012ameliorating,rui2017fast,sioulas2019hardware,yabuta2017relational}, and sort~\cite{gpu-terasort,gpu-sort2}.
Several full-fledged GPU-as-coprocessor database query engines have been proposed in recent years. Ocelot~\cite{ocelot} provides a hybrid analytical engine as an extension to MonetDB. YDB~\cite{yinyang} is a GPU-based data warehousing engine. 
Both systems used an operator-at-a-time model, where an
operator library containing GPU kernel implementations of common database
operators such as scans and joins is invoked on batches of tuples, running each
operator to completion before moving on to the next operator. 
Kernel fusion~\cite{wu2012kernel} attempted to hide in-efficiency associated with running multiple kernels for each query
like in the operator-at-a-time model. Kernel fusion fused operator kernels with producer-consumer dependency when possible to 
eliminate redundant data movement. As kernel fusion is applied as a post-processing step, it will miss opportunities where kernel 
configurations are incompatible (like the one in described in Section~\ref{sec:tile-model}).
HippogriffDB~\cite{hippogriffdb} used GPUs for large scale data warehousing where data resides on SSDs. HippogriffDB claims to achieve 100$\times$ speedup over MonetDB when the ratio of memory bandwidth of GPU to CPU is roughly 5$\times$. We have not been able to get the source code to compare against the system. 
More recently, HorseQC~\cite{funke2018pipelined} proposes pipelined data transfer between CPU and GPU to improve query runtime. 
As we show in the next section, using HorseQC ends up being slower than running the query efficiently directly on the CPU.

Commercial systems like Omnisci~\cite{omnisci}, Kinetica~\cite{kinetica}, and BlazingDB~\cite{blazingdb} aim to provide real-time analytical capabilities by using GPUs to store large parts (or all) of the working set. The setting used in this paper is similar to ones used by these systems.
Although these systems use a design similar to what we advocate, some have claimed 1000$\times$ performance improvement by using GPUs~\cite{mapd1000x} but have not published rigorous benchmarks against state-of-the art CPU or GPU databases, which is the primary aim of this paper.


\vspace{-.1in}
\subsection{Query Execution on CPU}

Database operators have been extensively optimized for modern processors. 
For joins, researchers have proposed using cache-conscious partitioning to improve hash join performance~\cite{manegold2000optimizing, blanas2011design, balkesen2013main, balkesen2013multi}. Schuh et al. summarized the approaches~\cite{schuh2016experimental}. 
For sort, Satish et al.~\cite{satish2010fast} and Wassenberg et al.~\cite{wassenberg2011engineering} introduced buffered partitioning for radix sort. Polychroniou et al.~\cite{polychroniou2014comprehensive} presented faster variants of radix sort that use SIMD instructions. 
Sompolski et al.~\cite{sompolski2011vectorization} showed that combination of vectorization and compilation can improve performance of project, selection, and hash join operators. Polychroniou et al.~\cite{polychroniou2015rethinking} presented efficient vectorized designs for selections, hash tables, and partitioning using SIMD gathers and scatters. Prashanth et al.~\cite{menon2017relaxed} extended the idea to generate machine code for full queries with SIMD operators. We use ideas from these works, mainly the works of Polychroniou et al.~\cite{polychroniou2014comprehensive,polychroniou2015rethinking} for our CPU implementations. 


C-Store~\cite{cstore} and MonetDB~\cite{monetdb} were among the first column-oriented engines, which formed the basis for analytical query processing. 
MonetDB X100~\cite{monetdbx100} introduced the idea of vectorized execution that was cache aware and reduced memory traffic. Hyper~\cite{hyper} introduced the push-based iteration and compiling queries into machine code using LLVM. Hyper was significantly faster than MonetDB and  brought query performance close to that of handwritten C code. We compare the performance of our CPU query implementations against \new{MonetDB} ~\cite{monetdb} and Hyper~\cite{hyper}.





\section{Our Approach}
\label{sec:our-approach}

\rev{In this section, we describe the tile-based execution model we use to execute queries on GPU efficiently.
We begin by showing why the coprocessor model used by past works is a suboptimal design and motivate why storing 
the working set directly on the GPU in a heterogeneous system (as done by all commercial systems) is a better approach. 
Through an example, we illustrate the unique challenges associated with running queries in a massively-parallel manner on the GPU.
We show how by treating the thread block as the basic execution unit with each thread block processing a tile of items (similar to vector-based processing on the CPU where each thread processes a vector of items a time) leads to good performance on the GPU. We call this approach the tile-based execution model. 
Finally, we show how this model can be expressed using a set of primitives where each primitive is a function which takes as input of set of tiles and outputs a set of tiles. We call these primitives  block-wide functions. We present \texttt{Crystal}, a library of block-wide functions that can be composed to create a full SQL query.}

\subsection{Failure of the Coprocessor Model}
\label{sec:failure-coop}

While past work has claimed speedups from using GPUs in the coprocessor model, there is no consensus among past
work about the performance improvement obtained from using GPUs, with reported improvements varying from $2\times$ to $100\times$. 

\begin{figure}[t]
\footnotesize
\begin{verbatim}
SELECT SUM(lo_extendedprice * lo_discount) AS revenue 
FROM lineorder
WHERE lo_quantity < 25 
AND lo_orderdate >= 19930101 AND lo_orderdate <= 19940101 
AND lo_discount >= 1 AND lo_discount <= 3;
\end{verbatim}
\vspace{-.1in}
\caption{Star Schema Benchmark Q1.1}
\label{fig:query11}
\vspace{-.1in}
\end{figure}
\normalsize

\begin{figure}[t]
  \centering
  \includegraphics[width=\columnwidth]{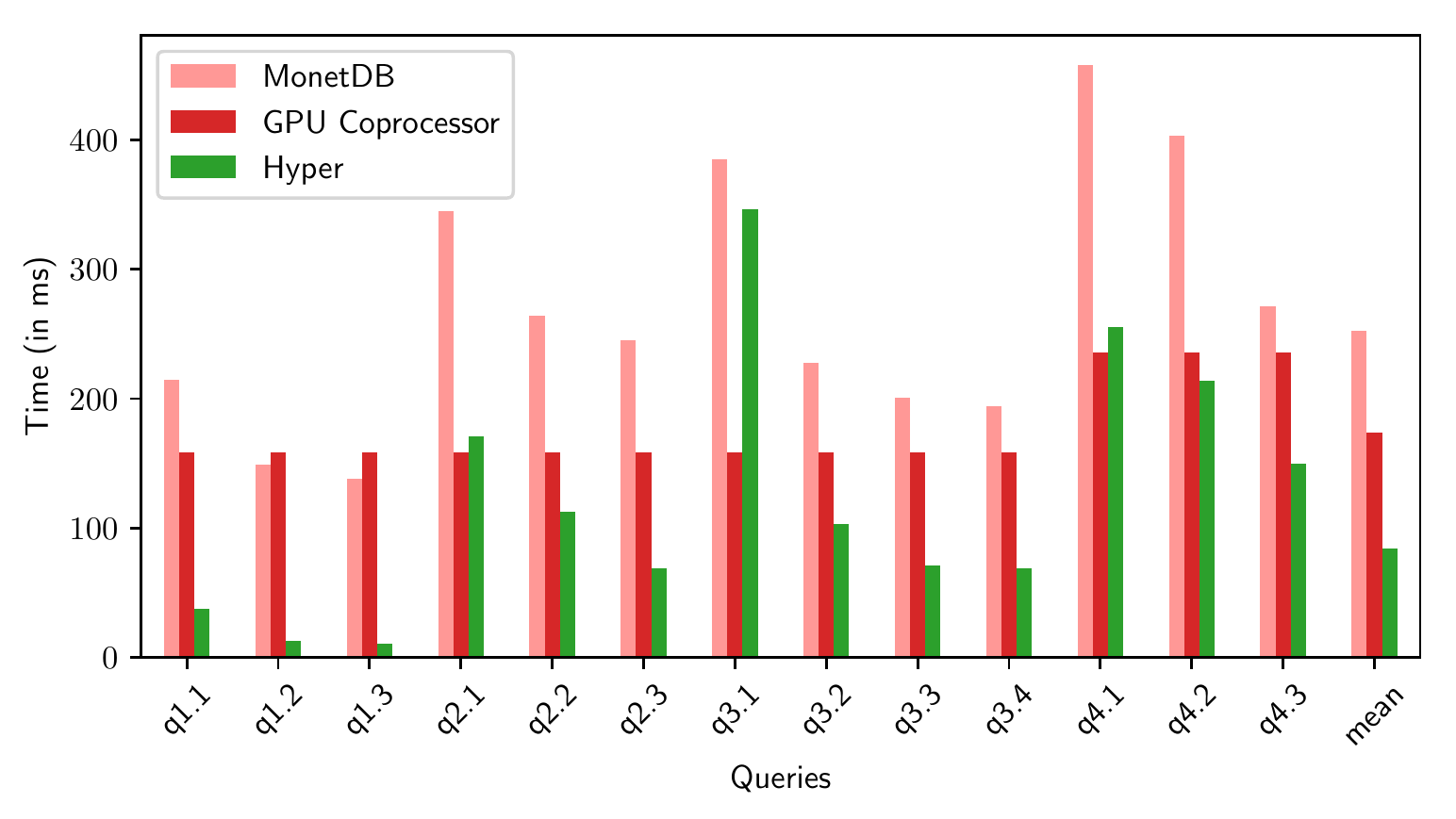}
  \vspace{-.3in}
  \caption{Evaluation on the Star Schema Benchmark}
  \label{fig:fullbench}
  \vspace{-.1in}
\end{figure}

Consider Q1.1 from the Star Schema Benchmark (SSB) shown in Figure~\ref{fig:query11}.
For simplicity, assume all column entries are 4-byte integers and $L$ is the number of entries in \texttt{lineorder}. An efficient implementation on a CPU will be able to generate the result using a single pass over the 4 data columns. The optimal CPU runtime ($R_C$) is upper bounded by $16L/B_c$ where $B_c$ is the CPU memory bandwidth. This is an upper bound because, if the predicates are selective, then we may be able to skip entire cache lines while accessing  the \texttt{lo\_extendedprice} column. In the coprocessor model, we have to ship 4 columns of data to GPU. Thus, the query runtime on the GPU ($R_G$) is lower bounded by $16L/B_p$ where $B_p$ is the PCIe bandwidth. The bound is hit if we are able to perfectly overlap the data transfer and query execution on GPU. However, since $B_c > B_p$ in modern CPU-GPU setups, $R_C < R_G$, i.e., running the query on CPU yields a lower runtime than the running query with a GPU coprocessor.

To show this empirically, we ran the entire SSB with scale factor 20 on an instance where CPU memory bandwidth is 54 GBps, GPU memory bandwidth is 880 GBps, and PCIe bandwidth is 12.8 GBps. The full workload details can be found in Section~\ref{sec:ssb-workload} and the full system details can be found in Table 2. We compare the performance of the \texttt{GPU Coprocessor} with two OLAP DBMSs: \texttt{MonetDB} and \texttt{Hyper}. Past work on using GPUs as a coprocessor mostly compared their performance against MonetDB~\cite{yinyang,gpudb2,hippogriffdb} which is known to be inefficient~\cite{hyper}. Figure~\ref{fig:fullbench} shows the results. On an average, \texttt{GPU Coprocessor} performs $1.5\times$ faster than \texttt{MonetDB} but it is $1.4\times$ slower than \texttt{Hyper}. For all queries, the query runtime in \texttt{GPU coprocessor} is bound by the PCIe transfer time. We conclude the reason past work was able to show performance improvement with a GPU coprocessor is because their optimized implementations were compared against inefficient baselines (e.g., MonetDB) on the CPU.

With the significant increase in GPU memory capacity, a natural question is how much faster a system that treats the GPU as the primary execution engine, rather than as an accelerator, can be.  We describe our architecture for such a system in the rest of this section.

\subsection{Tile-based Execution Model}
\label{sec:tile-model}

While a modern CPU can have dozens of cores, a modern GPU like Nvidia V100 can have 5000 cores. The vast increase in parallelism introduces some unique challenges for data processing. To illustrate this, consider running the following simple selection query as a micro-benchmark on both a CPU and a GPU:

\begin{lstlisting}[mathescape=true,basicstyle=\ttfamily]
Q0: SELECT y FROM R WHERE y > v;
\end{lstlisting}

On the CPU, the query can be efficiently executed as follows. The data is partitioned equally among the cores.  The goal is to write the results in parallel into a contiguous output array.
The system maintains a global atomic counter that acts as a cursor that tells each thread where to write the next result.
Each core processes its partition by iterating over the entries in the partition
one vector of entries at a time, where a vector is about 1000 entries (small enough to
fit in the L1 cache). Each core
makes a first pass over the first vector of entries to count the number of
entries that match the predicate $d$.
The thread increments the global counter by $d$ to allocate output space for the matching records, and then does a second pass over the vector to copy the matched entries into the output array in the allocated range of the output. Since the second pass reads data from L1 cache, the read is essentially free. The global atomic counter is a potential point of contention. However, note that each thread updates the counter once for every 1000+ entries and there are only around 32 threads running in parallel at any point. The counter ends up not being the bottleneck and the total runtime is approximately $\frac{D}{B_{C}} + \frac{D\sigma}{B_{C}}$ where $D$ is the size of the column, and $B_{C}$ is the memory bandwidth on the CPU.

\begin{figure}[t!]
  \centering
  \includegraphics[width=0.9\columnwidth]{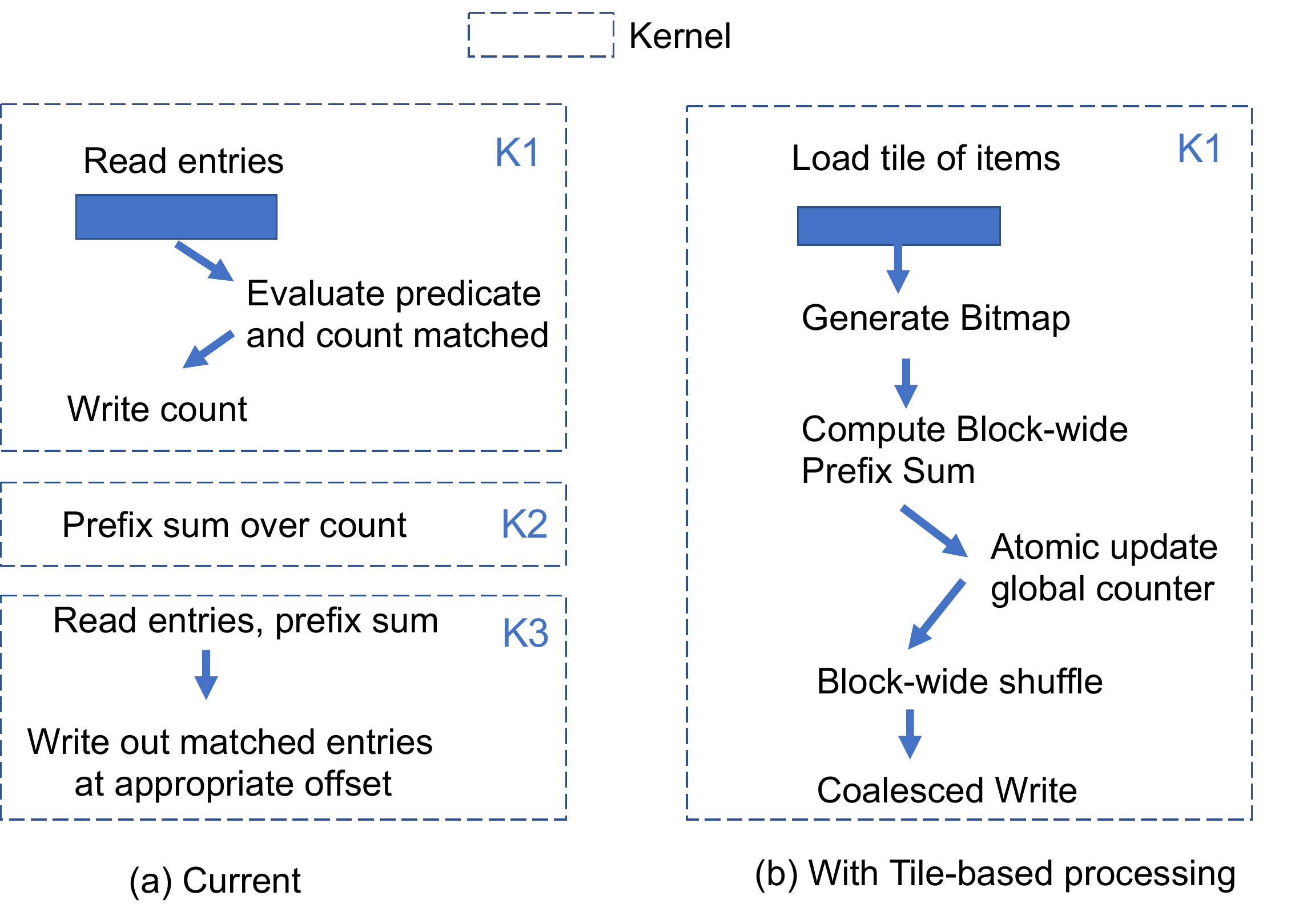}
  \vspace{-.2in}
  \caption{Running selection on GPU}
  \label{fig:tile}
  \vspace{-.2in}
\end{figure}

We could run the same plan on the GPU, partitioning the data up among the
thousands of threads. However, GPU threads have significantly fewer resources
per thread. On the Nvidia V100, each GPU thread can only store roughly 24 4-byte
entries in shared memory at full occupancy, with 5000 threads running in
parallel. Here, the global atomic counter ends up becoming the bottleneck as all
the threads will attempt to increment the counter to find the offset into the output array. To work around this, existing GPU-based database systems would execute this query in 3 steps as shown in Figure~\ref{fig:tile}(a). The first kernel $K_1$ would be launched across a large number of threads. In it, each thread would read in column entries
 in a strided fashion (interleaved by thread number) and evaluate the predicate to count
the number of entries matched. After processing all elements, the total number of entries matched per thread would be recorded in an array \texttt{count}, where \texttt{count[t]} is number of entries matched by thread \texttt{t}. The second kernel $K_2$ would use the \texttt{count} array to compute the prefix sum of the count and store this in another array
\texttt{pf}.  Recall that for an array $A$ of $k$ elements, the prefix sum $p_A$ is a $k$ element array where $p_A[j] = \sum_{i=0}^{j-1}{A_j}$. Thus, the $i^{th}$ entry in {\tt pf} indicates the offset at which the $i$th thread should write its matched results to in the output array \texttt{o}. Databases used an optimized routine from a CUDA library like Thrust~\cite{thrust} to run it efficiently in parallel. 
The third kernel $K_3$ would then read in the input column again; here the \new{$i^{th}$}
thread again scans the \new{$i^{th}$} stride of the input, using {\tt pf}$[i]$ to
determine where to write the satisfying records.  Each thread also maintains a
local counter $c_i$, initially set to 0. Specifically for each satisfying entry,
thread $i$ writes it to \texttt{pf}$[i]+c_i$ and then increments $c_i$. In the
end, \texttt{o[pf[t]] ... o[pf[t+1] - 1]} will contain the matched entries of thread \texttt{t}.

The above approach shifts the task of finding offsets into the output array to an optimized prefix sum kernel whose runtime is a function of $T$ (where $T$ is the number of threads $(T << n)$), instead of finding it inline using atomic updates to a counter. As a result, the approach ends up being significantly faster than the naive translation of the CPU approach to the GPU. However, there are a number of issues with this approach. First, it reads the input column from global memory twice, compared to doing it just once with the CPU approach. It also reads/writes to intermediate structures \texttt{count} and \texttt{pf}. Finally, each thread writes to a different location in the output array resulting in random writes. To address these issues, we introduce the \textbf{Tile-based execution model}.

\begin{figure}[t!]
  \centering
  \includegraphics[width=0.8\columnwidth]{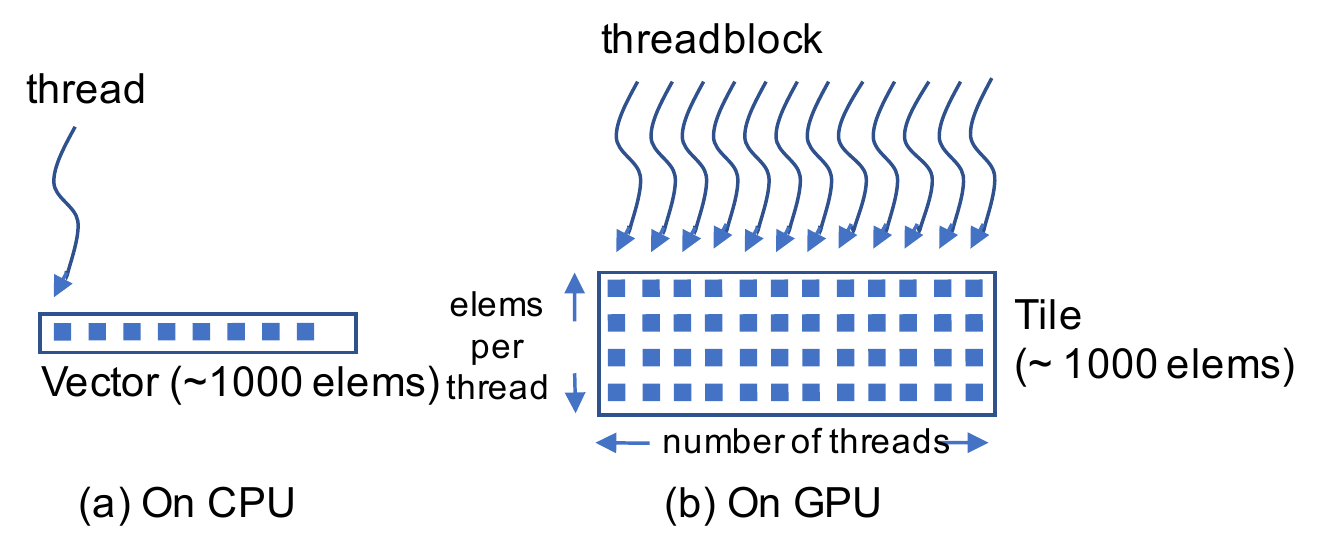}
  \vspace{-.1in}
  \caption{Vector-based to Tile-based execution models.}
  \label{fig:tile-intuition}
  \vspace{-.1in}
\end{figure}

Tile-based processing extends the vector-based processing on CPU where each thread processes a vector at a time to the GPU. Figure~\ref{fig:tile-intuition} illustrates the model. As discussed earlier in Section~\ref{sec:gpu-memory-hierarchy}, threads on the GPU are grouped into thread blocks. Threads within a thread block can communicate through shared memory and can synchronize through barriers. Hence, even though a single thread on the GPU at full occupancy can hold only up to 24 integers in shared memory, a single thread block can hold a significantly larger group of elements collectively between them in shared memory. We call this unit a \textbf{Tile}. 
In the Tile-based execution model, instead of viewing each thread as an independent execution unit, we view a thread block as the basic execution unit with each thread block processing a tile of entries at a time.  One key advantage of this approach is that after a tile is loaded into shared memory, subsequent passes over the tile will be read directly from shared memory and not from global memory, avoiding the second pass through global memory described  in the implementation above.

\begin{figure}[t!]
  \centering
  \includegraphics[width=0.9\columnwidth]{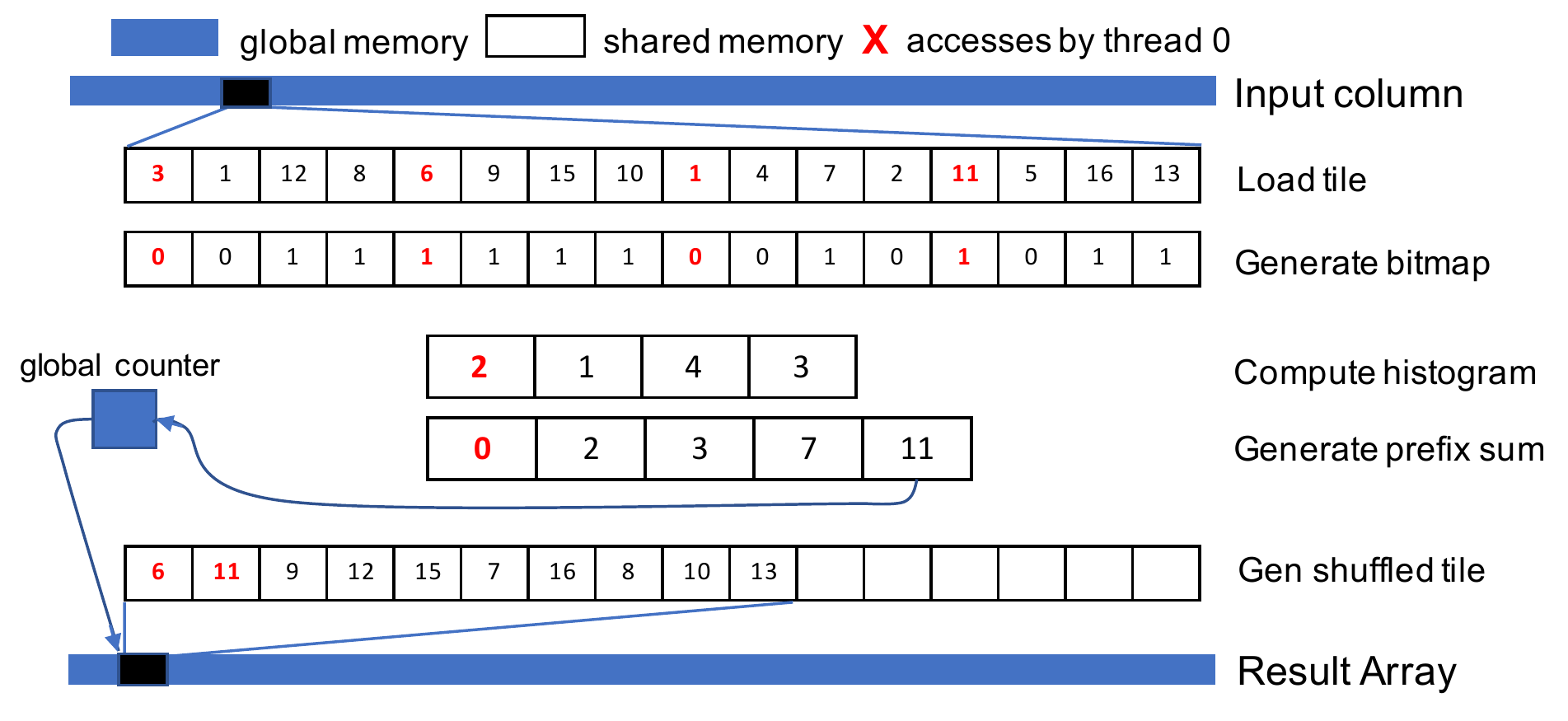}
  \vspace{-.1in}
  \caption{Query Q0 Kernel running y $>$ 5 with tile size 16 and thread block size 4}
  \label{fig:tile-example}
  \vspace{-.1in}
\end{figure}

\rev{
Figure~\ref{fig:tile}(b) shows how selection is implemented using the
tile-based model. The entire query is implemented as a single kernel
instead of three. Figure~\ref{fig:tile-example} shows a sample execution with a
tile of size 16 and a thread block of 4 threads for the predicate $y > 5$. Note
that this is just for illustration, as most modern GPUs would use a thread block
size that is a multiple of 32 (the warp size) and the number of elements loaded
would be 4--16 times the size of the thread block. We
start by initializing the global counter to 0. The kernel loads a tile of items
from global memory into the shared memory. The threads then apply the predicate
on all the items in parallel to generate a bitmap. For example, thread 0
evaluates the predicate for elements 0,4,8,12 (shown in red). Each thread then
counts the number of entries matched per thread to generate a histogram. The
thread block co-operatively computes the prefix sum over the histogram to find
the offset each thread writes to in shared memory. In the example, threads
0,1,2,3 match 2,1,4,3 entries respectively. The prefix sum entries 0,2,3,7 tell
us thread 0 should write its matched entries to output at index 0, thread
1 should write starting at index 2, etc. We increment a global counter atomically by total number of matched entires to find the
offset at which the thread block should write in the output array. The shuffle
step uses the bitmap and the prefix sum to create a contiguous array of matched
entries in shared memory. The final write step copies the contiguous entries
from shared memory to global memory at the right offset.
}

By treating the thread block as an execution unit, we reduce the number atomic
updates of the global counter by a factor of size of tile $T$. 
The kernel also makes a single pass
over the input column with the \texttt{Gen Shuffled Tile} ensuring that the
final write to the output array is coalesced, solving both problems associated
with approach used in previous GPU databases. 

\new{The general concept of the tile-based executing model i.e., dividing data into tiles and mapping threadblocks to tiles has been used in other domains like image processing~\cite{holewinski2012high} and high performance computing~\cite{thrust}. However, to the best of our knowledge this is the first work that uses it for database operations. In the next section, we present \texttt{Crystal}, a library of data processing primitives that can be composed together to implement SQL queries on the GPU.}

\begin{figure}[t]
  \centering
  \includegraphics[width=\columnwidth]{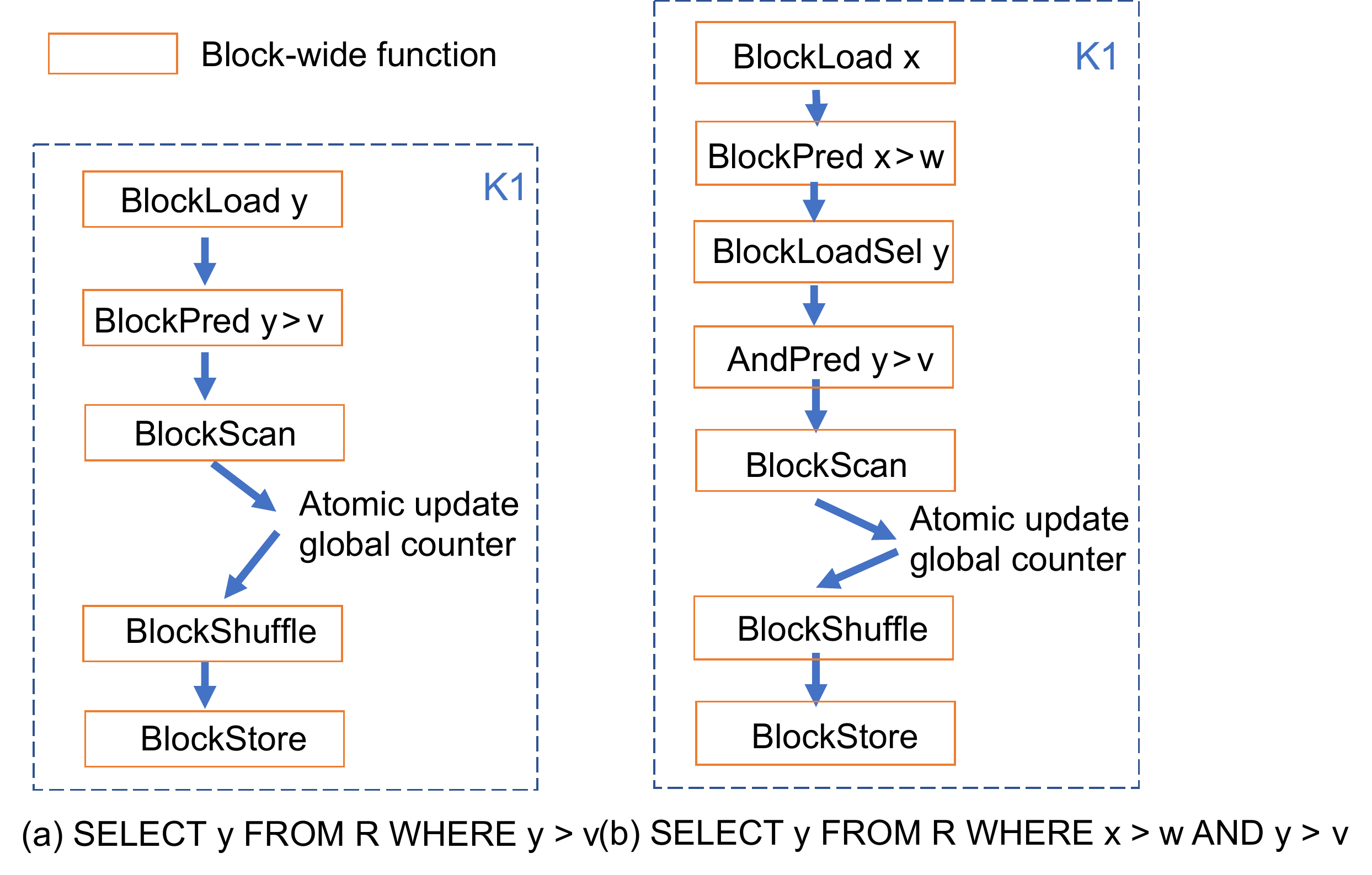}
  \vspace{-.2in}
  \caption{Implementing queries using Crystal}
  \label{fig:tile-extension}
  \vspace{-.1in}
\end{figure}




\subsection{Crystal Library}

\begin{table*}[t!]
\centering
\footnotesize
\begin{tabular}{|l|p{14.5cm}|}
\hline
Primitive & Description \\
\hline
\texttt{BlockLoad} & Copies a tile of items from global memory to shared memory. Uses vector instructions to load full tiles. \\
\texttt{BlockLoadSel} & Selectively load a tile of items from global memory to shared memory based on a bitmap. \\
\texttt{BlockStore} & Copies a tile of items in shared memory to device memory. \\
\texttt{BlockPred} & Applies a predicate to a tile of items and stores the result in a bitmap array. \\
\texttt{BlockScan} & Co-operatively computes prefix sum across the block. Also returns sum of all entries. \\
\texttt{BlockShuffle} & Uses the thread offsets along with a bitmap to locally rearrange a tile to create a contiguous array of matched entries. \\
\texttt{BlockLookup} & Returns matching entries from a hash table for a tile of keys. \\
\texttt{BlockAggregate} & Uses hierarchical reduction to compute local aggregate for a tile of items. \\
\hline
\end{tabular}
\caption{List of block-wide functions}
\vspace{-.2in}
\label{table:list-func}
\end{table*}

\new{The kernel structure in Figure~\ref{fig:tile-example} contains a series of
steps where each is a function that takes as input a set of tiles, and
outputs a set of tiles. We call these primitives \textit{block-wide functions}.
A block-wide function is a device function\footnote{Device functions are functions that can be called from kernels on the GPU} that takes in a set of tiles as input,
performs a specific task, and outputs a set of tiles. Instead of reimplementing these 
block-wide functions for each query, which would involve repetition of non-trivial functions, 
we developed a library called \textbf{\texttt{Crystal}}.}

\begin{figure}[t]
\includegraphics[width=\columnwidth]{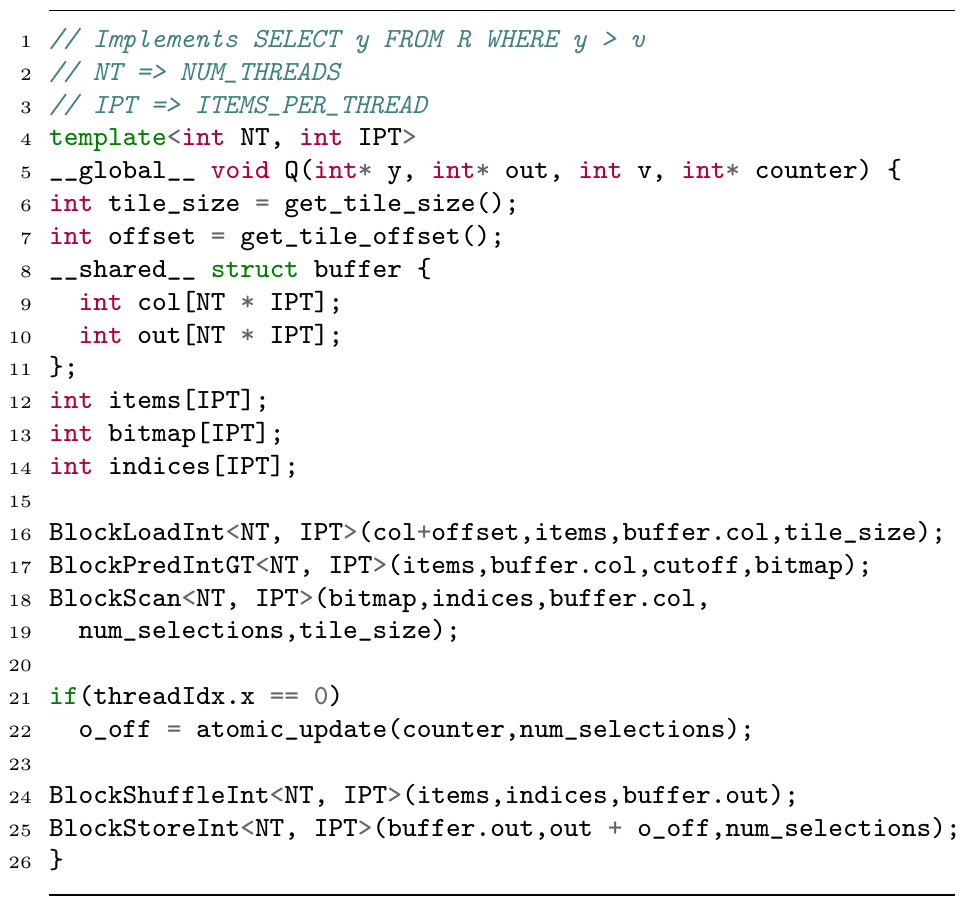}
\caption{Query Q0 Kernel Implemented with \texttt{Crystal}}
\vspace{-.1in}
\label{fig:qcrystal}
\end{figure}

\new{\texttt{Crystal}\footnote{The source code of the Crystal library is available at \texttt{\href{https://github.com/anilshanbhag/crystal}{https://github.com/anilshanbhag/crystal}}} is a library of templated CUDA device functions that implement the full set of 
primitives necessary for executing typical analytic SQL SPJA analytical queries. 
Figure~\ref{fig:tile-extension}(a) shows an sketch of the simple selection query implemented using block-wide functions.
Figure~\ref{fig:qcrystal} shows the query kernel of the same query implemented with \texttt{Crystal}.
We use this example to illustrate the key features of \texttt{Crystal}.
The input tile is loaded from the global memory into the thread block using \texttt{BlockLoad}.
\texttt{BlockLoad} internally uses vector instructions when loading a full tile and for the tail of the input array that may not
form a perfect tile, it is loaded in a striped fashion element-at-a-time. 
\texttt{BlockPred} applies the predicate to generate the bitmap. A key optimization that we do in \texttt{Crystal} is
instead of storing the tile in shared memory, in cases where the array indices are statically known before hand, we choose
to use registers to store the values. In this case, \texttt{items} (which contains entries loaded from the column) and \texttt{bitmap} 
are stored in registers. Hence, in addition to 24 4-byte values that a thread can store in shared memory, this technique allows us to use roughly equal amount of registers available to store data items.
Next we use \texttt{BlockScan} to compute the prefix sum. \texttt{BlockScan} internally
implements a hierarchical block-wide parallel prefix-sum approach~\cite{harris2007parallel}. This involves threads accessing bitmap entries of other threads --- for this we load \texttt{bitmap} into shared memory, reusing \texttt{buffer.col} shared memory buffer used for loading the input column. Shared memory is order of magnitude faster than global memory, hence loads and stores to shared memory in this case do not impact performance.
After atomic update to find offset in output array, \texttt{BlockShuffle} is used to reorder the array 
and finally we use \texttt{BlockStore} to write to output array. The code skips some minor details like when the atomic update happens, 
since it is executed on thread 0, the global offset needs to be communicated back to other threads through shared memory.
}

\new{In addition to allowing users to write high performance kernel code that as we show later can saturate memory bandwidth,
there are two usability advantages of using \texttt{Crystal}:}
\vspace{-0.05in}
\begin{itemize} [leftmargin=*]
\item \new{\textbf{Modularity:} Block-wide functions in \texttt{Crystal} make it easy to use non-trivial functions
and reduce boilerplate code. For example, \texttt{BlockScan}, \texttt{BlockLoad}, \texttt{BlockAggregate} each encapsulate 10's to 100's of lines of code.
For the selection query example, \texttt{Crystal} reduces lines of code from more than 300 to less than 30.}
\item \new{\textbf{Extensibility:} Block-wide functions makes it is fairly easy 
to implement query kernels of larger queries. Figure~\ref{fig:tile-extension}(b) shows the
implementation of a selection query with two predicates. Ordinary CUDA code can be used along with \texttt{Crystal} functions.}
\end{itemize}
\vspace{-0.05in}

\new{\texttt{Crystal} supports loading partial tiles like in Figure~\ref{fig:tile-extension}(b). If a selection or join filters entries, we use 
\texttt{BlockLoadSel} to load items that matched the previous selections based on a bitmap. In this case, the thread block internally allocate space for the entire tile, however, only matched entries are loaded from global memory.}
\new{Table~\ref{table:list-func} briefly describes the
block-wide functions currently implemented in the library.}

\new{To evaluate Crystal, we look at two microbenchmarks:}

\begin{figure}[t]
  \centering
  \includegraphics[width=0.9\columnwidth]{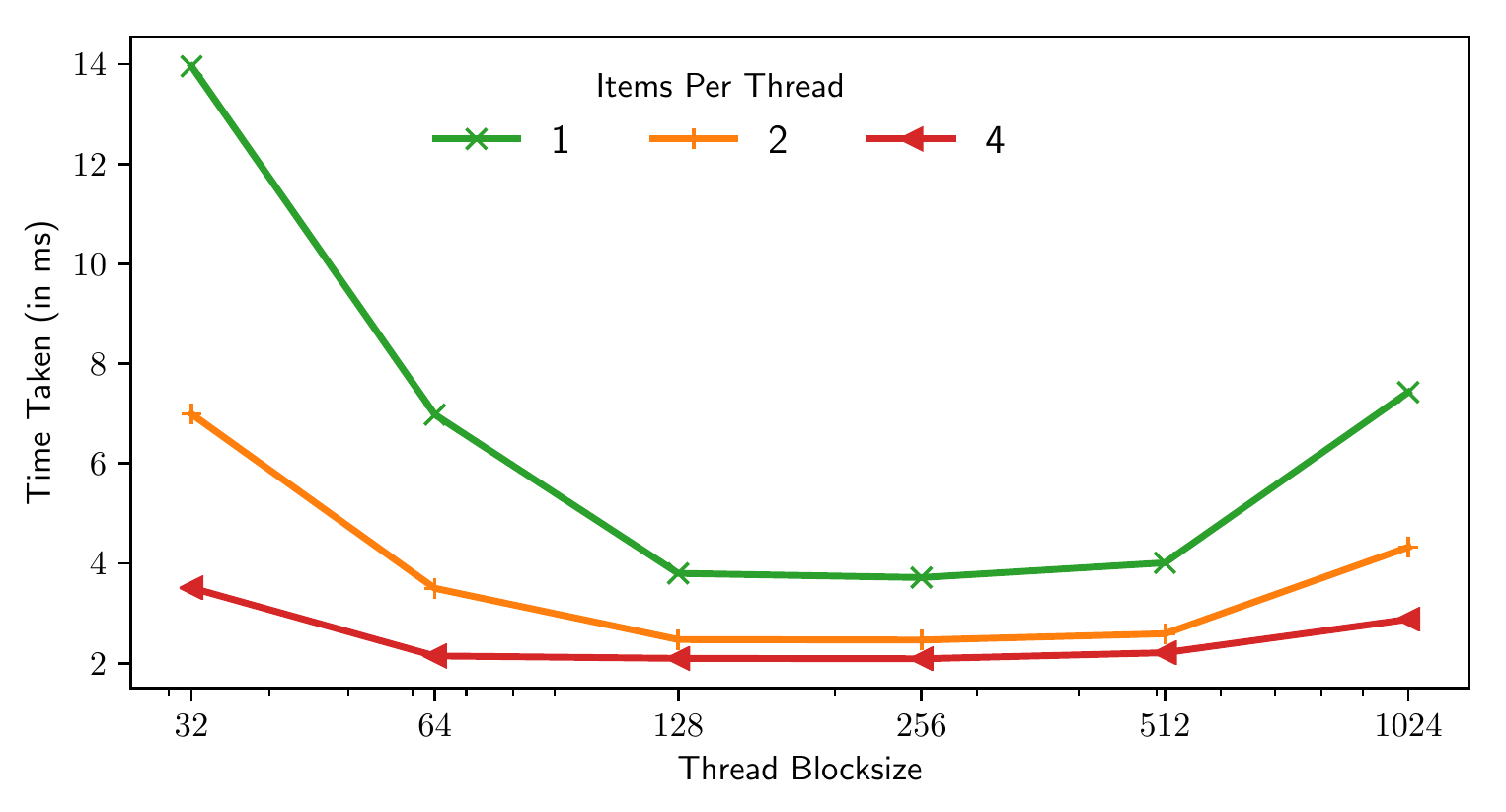}
  \vspace{-.2in}
  \caption{Q0 performance with varying tile sizes}
  \label{fig:tile-sizes}
  \vspace{-.1in}
\end{figure}

\noindent \new{1) We evaluate the selection query $Q0$ with size of input array as $2^{29}$ and selectivity is $0.5$. We vary the tile sizes. We vary the thread block sizes from 32 to 1024 in multiples of 2. We have three choices for the number of items per thread: 1,2,4. Figure~\ref{fig:tile-sizes} shows the results. 
As we increase the thread block size, the number of global atomic updates done reduces and hence the runtime improves until the thread block size approaches 512 after which it deteriorates. Each streaming multiprocessor on the GPU holds maximum of 2048 threads, hence, having large thread blocks reduces number of independent thread blocks. This affects utilization particularly when thread blocks are using synchronization heavily. Having 4 items per thread allows to effectively load the entire block using vector instructions. With 2 items per thread, there is reduced benefit for vectorization as half the threads are empty. With 1 item per thread there is no benefit. The best performance is seen with thread block size of 128/256 and items per thread equal to 4. In these cases, as we show later in Section~\ref{sec:select-op} saturate memory bandwidth and hence achieve optimal performance. }

\noindent \new{2) We evaluated the selection query $Q0$ using two approaches: independent threads approach (Figure~\ref{fig:tile}(a)) and using \texttt{Crystal} (Figure~\ref{fig:tile}(b)). The number of entries in the input array is $2^{29}$ and selectivity is $0.5$. The runtime with the independent threads approach is 19ms compared to just 2.1ms when using \texttt{Crystal}. Almost all of the performance improvement is from avoiding atomic contention and being able to reorder matched entries to write in a coalesced manner}.

\new{Across all of the workloads we evaluated, we found that using thread block size 128 with items per thread equal to 4 is indeed the best performing tile configuration. In the rest of the paper, we use this configuration for all implementations using \texttt{Crystal}. 
All the implementations with \texttt{Crystal} are implemented in CUDA C++. Since \texttt{Crystal}'s block-wide functions are standard device functions, they can also called directly from LLVM IR}.

\new{In the next section, we show how to use these block-wide functions to build
efficient operators on a GPU and compare their performance to equivalent CPU implementations.}

\section{Operators on GPU vs CPU}
\label{sec:opcomp}

In order to understand the true nature of performance difference of queries on GPU vs. CPU, it is important to understand the performance difference of individual query operators. In this section, we compare the performance of basic operators: project, select, and hash join on GPU  and  CPU with the goal of understanding how the ratio of runtime on GPU to runtime on CPU compares to the bandwidth ratio of the two devices. We use block-wide functions from \texttt{Crystal} to implement the operators on GPU  and use equivalent state-of-the-art implementations on CPU. We also present a model for each of the operators assuming the operator saturates memory bandwidth and show that in most cases the operators indeed achieve these limits. We use the model to explain the performance difference between CPU and GPU. For the micro-benchmarks, we use a setup where GPU memory bandwidth is 880GBps and CPU memory bandwidth is 54GBps, resulting in a bandwidth ratio of $16.2$ (see Section~\ref{sec:evaluation} for system details). In all cases, we assume that the data is already in the respective device's memory.

\subsection{Project}

We consider two forms of projection queries: one that computes a linear combination of columns (Q1) and one involving user defined function (Q2) as shown below:
\vspace{-0.05in}
\begin{lstlisting}[mathescape=true,basicstyle=\ttfamily]
Q1: SELECT a$x_1$ + b$x_2$ FROM R;
Q2: SELECT $\sigma$(a$x_1$ + b$x_2$) FROM R;
\end{lstlisting}
\vspace{-0.05in}
\noindent where $x_1$ and $x_2$ are 4-byte floating point values. The number of entries in the input array is $2^{29}$. $\sigma$ is the sigmoid function (i.e., $\sigma(x) = \frac{1}{1+e^{-x}}$) which
can represent the output of a logistic regression model. Note that $Q1$ consists of basic arithmetic and will certainly be bandwidth bound. $Q2$ is representative of the most complicated projection we will likely see in any SQL query.

On the CPU side, we implement two variants: \texttt{CPU} and \texttt{CPU-Opt}. \texttt{CPU} uses a multi-threaded projection where each thread works on a partition of the input data. 
\texttt{CPU-Opt} extends \texttt{CPU} with two extra optimizations: (1) non-temporal writes and (2) SIMD instructions. 
Non-temporal writes are write instructions that bypass higher cache levels and write out an entire cache line to main memory without first loading it to caches. 
SIMD instructions can further improve performance.
With a single AVX2 instruction, for example, a modern x86 system can add, subtract, multiply, or divide a group of 8 4-byte floating point numbers, thereby improving the computation power and memory bandwidth utilization. 

On the GPU side, we implement a single kernel that does two \texttt{BlockLoad}'s to load the tiles of the respective columns, computes
the projection and does a \texttt{BlockStore} to store it in the result array.

\vspace{.05in}
\noindent\textbf{Model:} Assuming the queries can saturate the memory bandwidth, the expected runtime of Q1 and Q2 is 

\vspace{-0.12in}
\[ \textit{runtime} = \frac{2\times4\times N}{B_r} + \frac{4\times N}{B_w} \]
\vspace{-0.12in}

\noindent where $N$ is the number of entries in the input array and $B_r$ and $B_w$ are the read and write memory bandwidth, respectively. The first term of the formula models the runtime for loading columns $x_1$ and $x_2$, each containing 4-byte floating point numbers. 
The second term models the runtime for writing the result column back to memory, which also contains 4-byte floating point numbers. 
Note that this formula works for both CPU and GPU, by plugging in the corresponding memory bandwidth numbers. 

\begin{figure}[t]
  \centering
  \includegraphics[width=\columnwidth]{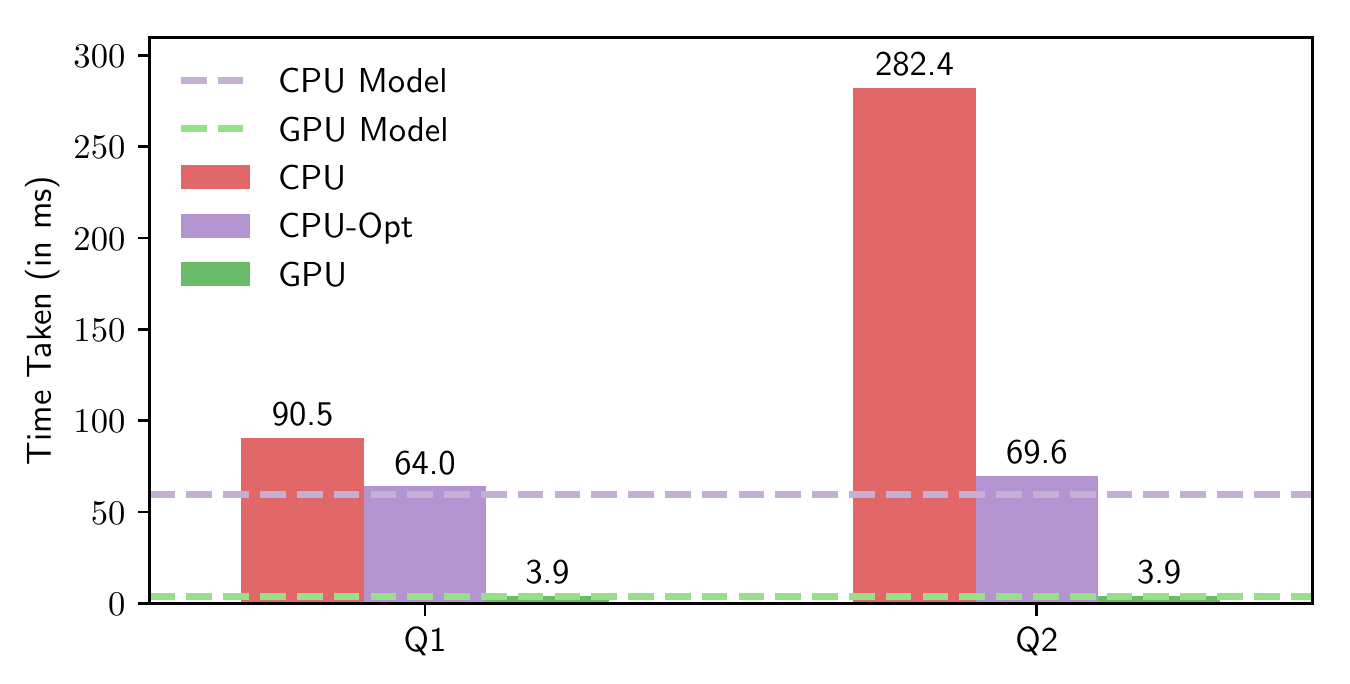}
  \vspace{-.3in}
  \caption{Project microbenchmark}
  \label{fig:projectres}
  \vspace{-.1in}
\end{figure}

\vspace{.05in}
\noindent\textbf{Performance Evaluation:} 
Figure~\ref{fig:projectres} shows the runtime of queries Q1 and Q2 on both CPU and GPU (shown as bars) as well as the predicted runtime based on the model (shown as dashed lines). 
The performance of Q1 on both CPU and GPU is memory-bandwidth bound. 
\texttt{CPU-Opt} performs better than \texttt{CPU} due
to the increased memory bandwidth efficiency. 
\texttt{GPU} performs substantially better than both CPU implementations due to its much higher memory bandwidth. 
The ratio of runtime of \texttt{CPU-Opt} to \texttt{GPU}
is $16.56$ which is close to the bandwidth ratio of $16.2$. The minor difference is because read bandwidth is slightly lower than write bandwidth on the CPU and the workload
has a read:write ratio of 2:1.

A simple multi-threaded implementation of Q2 (i.e., \texttt{CPU}) does not saturate memory
bandwidth and is compute bound. 
After using the SIMD instructions (i.e, \texttt{CPU-Opt}), performance improves significantly and the system is close to memory bandwidth bound. 
The ratio of runtime of \texttt{CPU-Opt} to \texttt{GPU} for Q2 is $17.95$. 
This shows that even for fairly complex projections, good implementations on modern CPUs are able to saturate memory bandwidth. 
GPUs do significantly better than CPUs due to their high memory bandwidth, with the performance gain equal to the bandwidth ratio.


\subsection{Select} 
\label{sec:select-op}

We now turn our attention to evaluating selections, also called selection scans. Selection scans have
re-emerged for main-memory query execution and are replacing tradition unclustered indexes in modern OLAP
DBMS~\cite{raman2013db2}. We use the following micro-benchmark to evaluate selections:

\vspace{-0.05in}
\begin{lstlisting}[mathescape=true,basicstyle=\ttfamily]
Q3: SELECT y FROM R WHERE y < v;
\end{lstlisting}
\vspace{-0.05in}

\noindent where $y$ and $v$ are both 4-byte floating point values. The size of input array is $2^{29}$. We vary the selectivity of the predicate
from 0 to 1 in steps of 0.1.

To evaluate the above query on a multi-core CPU, we use the CPU implementation described earlier in Section~\ref{sec:tile-model}.
We evaluate three variants. The ``naive'' branching
implementation (\texttt{CPU If}) implements the selection using an if-statement, as shown in Figure~\ref{fig:selscan-example}(a). 
The main problem with the branching 
implementation is the penalty for branch mispredictions. If the selectivity of the condition is neither too high 
nor too low, the CPU branch predictor is unable to predict the branch outcome. This leads to pipeline stalls that hinder performance. 
Previous work has shown that the branch misprediction penalty can be avoided by using branch-free 
\textit{predication} technique~\cite{ross2004selection}. 
Figure~\ref{fig:selscan-example}(b) illustrates the predication approach. Predication transforms the branch (control dependency) into a data dependency. \texttt{CPU Pred} implements selection scan with predication. More recently, vectorized
selection scans have been shown to improve on \texttt{CPU Pred} by using selective stores to buffer entries that satisfy selection predicates
and writing out entries using streaming stores~\cite{polychroniou2015rethinking}. \texttt{CPU SIMDPred} implements this approach.

\begin{figure}
\centering
\begin{subfigure}[b]{0.45\columnwidth}
\begin{verbatim}
for each y in R:
  if y > v:
    output[i++] = v
\end{verbatim}
\caption{With branching}
\end{subfigure}
    ~ 
\begin{subfigure}[b]{0.45\columnwidth}
\begin{verbatim}
for each y in R:
  output[i] = y
  i += (y > v)
\end{verbatim}
\caption{With predication}
\end{subfigure}
\caption{Implementing selection scan}
\label{fig:selscan-example}
\vspace{-.1in}
\end{figure}

On the GPU, the query is implemented as a single kernel as described  in Section~\ref{sec:tile-model} and as
shown in Figure~\ref{fig:tile}(b). We implement two variants: \texttt{GPU If} implements
the selection using an if-statement and \texttt{GPU Pred} implements it using predication.

\vspace{.05in}
\noindent\textbf{Model:} The entire input array is read and only the matched entries are written to the output array. Assuming the implementations 
can write out the matched entries efficiently and saturate memory bandwidth, the expected runtime is: 

\vspace{-0.12in}
\[ \textit{runtime} = \frac{4 \times N}{B_r} + \frac{4\times \sigma\times N}{B_w} \]
\vspace{-0.12in}

\noindent where $N$ is the number of entries in the input array, $B_r$ and $B_w$ are the read and write  bandwidth of the respective device, and $\sigma$ is the predicate selectivity.

\begin{figure}[t]
  \centering
  \includegraphics[width=\columnwidth]{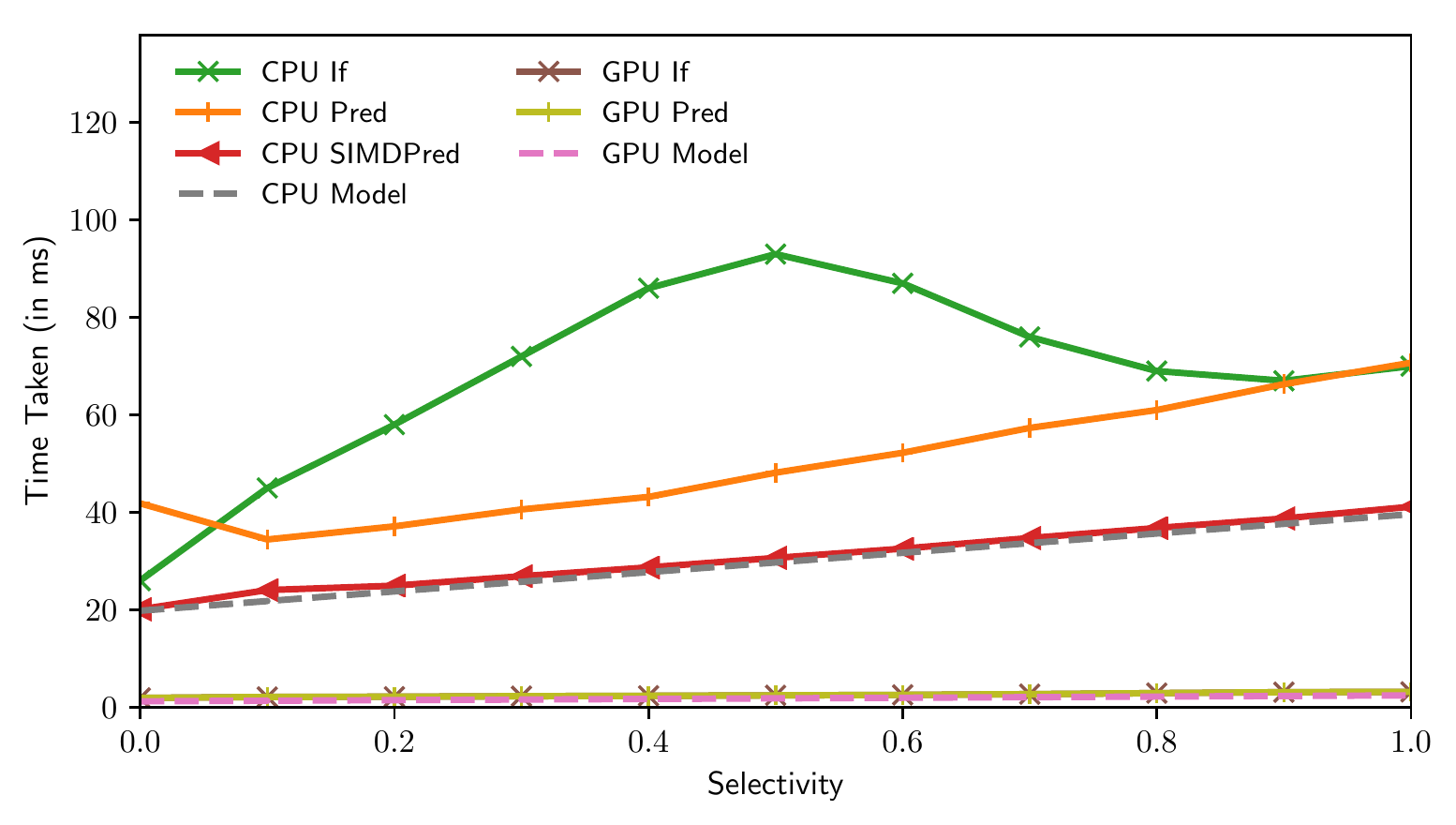}
  \vspace{-.3in}
  \caption{Select Microbenchmark}
  \label{fig:select}
  \vspace{-.1in}
\end{figure}

\vspace{.05in}
\noindent\textbf{Performance Evaluation:}
Figure~\ref{fig:select} shows the runtime of the three algorithms on CPU, two algorithms on GPU, and the performance models. \texttt{CPU Pred} does better 
than \texttt{CPU If} at all selectivities except 0 (at 0, \texttt{CPU If} does no writes). Across the range, \texttt{CPU SIMDPred}
does better than the two scalar implementations. On GPU, there is no performance difference between \texttt{GPU Pred} and \texttt{GPU If} --- A single branch misprediction does not impact performance on the GPU. Both \texttt{CPU SIMDPred} and \texttt{GPU If/Pred} closely 
track their respective theoretical models which assume saturation of memory bandwidth. 
The average runtime ratio of CPU-to-GPU is $15.8$
which is close to the bandwidth ratio $16.2$. This shows that with efficient implementations, CPU implementations saturate memory 
bandwidth for selections and the gain of GPU over CPU is equal to the bandwidth ratio. 



\subsection{Hash Join}
\label{sec:hash-join-sec}

\rev{Hash join is the most popular algorithm used for executing joins in a database. Hash joins have been extensively studied in the database literature, with many different hash join algorithms proposed for both CPUs and GPUs~\cite{balkesen2013main,blanas2011design,lang2015massively,balkesen2013multi,chen2007improving,gpu-join1}. 
The most commonly used hash join algorithm is the no partitioning join, which uses a non-partitioned global hash table. 
The algorithm consists of two phases: in the \textit{build phase}, the tuples in one relation (typically the smaller relation) are used to populate the hash table in parallel; in the \textit{probe phase}, the tuples in the other relation are used to probe the hash table for matches in parallel. For our microbenchmark, we focus on the following join query:}

\vspace{-0.05in}
\begin{lstlisting}[mathescape=true,basicstyle=\ttfamily]
Q4: SELECT SUM(A.v + B.v) AS checksum
    FROM A,B WHERE A.k = B.k
\end{lstlisting}
\vspace{-0.05in}

\noindent where each table A and B consists of two 4-byte integer columns $k,v$. The two tables are joined on key $k$. 
We keep the size of the probe table fixed at 256 million tuples, totaling ~2 GB of raw data. We use a hash table with 50\% fill rate. We vary the size of the build table such that it produces a hash table of the desired size in the experiment. We vary the size of the hash table from 8KB to 1GB. The microbenchmark is the same as what past works use~\cite{blanas2011design,balkesen2013main,balkesen2013multi,schuh2016experimental}.

In this section, we mainly focus on the probe phase which forms the majority of the total runtime. 
We discuss briefly the difference in execution with respect to build time at the end of the section. 
There are many hash table variants, in this section we focus on linear probing due to its simplicity and regular memory access pattern; our conclusions, however, should apply equally well to other probing approaches. 
Linear probing is an open addressing scheme that, to either insert an entry or terminate the search, traverses the table linearly until an empty bucket is found. 
The hash table is simply an array of slots with each slot containing a key and a payload but no pointers.

On the CPU side, we implemented three variants of linear probing. 
(1) \texttt{CPU Scalar} implements a scalar tuple-at-a-time join. The probing table is partitioned equally among the threads. Each thread iterates over its entries and for each entry probes the hash table to find a matching entry. On finding a match, it adds $A.v + B.v$ to its local sum. At the end, we add the local sum to the global sum using atomic instructions. 
(2) \texttt{CPU SIMD} implements vertical vectorized probing in a hash table~\cite{polychroniou2015rethinking}. The key idea in vertical vectorization is to process a different key per SIMD lane and use gathers to access the hash table. Assuming $W$ vector lanes, we process $W$ different input keys on each loop iteration. In every round, for the set of keys that have found their matches, we calculate their sum, add it to a local sum, and reload those SIMD lanes with new keys. 
(3) Finally, \texttt{CPU Prefetch} adds group prefetching to \texttt{CPU Scalar}~\cite{chen2007improving}. For each loop iteration, software prefetching instructions are inserted to load the hash table entry that will be accessed a certain number of loop iterations ahead. The goal is to better hide memory latency at the cost of increased number of instructions. 

On the GPU side, we implemented the join as follows. We load in a tile of keys and payloads from the probe side using \texttt{BlockLoad}; the threads iterate over each tile independently to find matching entries from the hash table. Each thread maintains a local sum of entries processed. After processing all entries in a tile, we use \texttt{BlockAgg} to aggregate the local sums within a thread block into a single value and increment a global sum with it.

\vspace{.05in}
\noindent\textbf{Model:} The probe phase involves making random accesses to the hash table to find the matching tuple from the build side. Every random access to memory ends up reading an entire cache line. However, if the size of hash table is small enough such that it can be cached, 
then random accesses no longer hit main memory and performance improves significantly. We model the runtime as follows:

\noindent 1) \new{If the hash table size is smaller than the size of the $K^{th}$ level cache, we expect the runtime to be:
\vspace{-0.05in}
\begin{align*}
runtime = max(\frac{4 \times 2 \times |P|}{B_r}, (1 - \pi_{K-1})(\frac{|P| \times C}{B_K}))
\end{align*}
\noindent where $|P|$ is the cardinality of the probe table, $B_r$ is the read bandwidth from device memory, $C$ is the cache line size accessed on probe, $B_K$ is the bandwidth of level $K$ cache in which hash table fits and $\pi_{K-1}$ is the probability of an access hitting a $K-1$ level cache. 
The first term is the time taken to scan the probe table from device memory. The second term is the time for probing the hash table. Note that each probe accesses an entire cache line. If the size of level $K$ cache is $S_K$ and size of the hash table is H, we define cache hit ratio $\pi_K = min(S_K/H, 1)$.
The total runtime will be bounded by either the device memory bandwidth or the cache bandwidth. Hence, the runtime is the maximum of the two terms. }

\noindent 2) If the hash table size is larger than the size of the last level cache, we expect the runtime to be:
\vspace{-0.05in}
\begin{align*}
runtime = \frac{4 \times 2 \times |P|}{B_r} + (1 - \pi)(\frac{|P| \times C}{B_r})
\end{align*}
where $\pi$ is the probability that the accessed cache line is the last level cache.

\begin{figure}[t]
  \centering
  \includegraphics[width=\columnwidth]{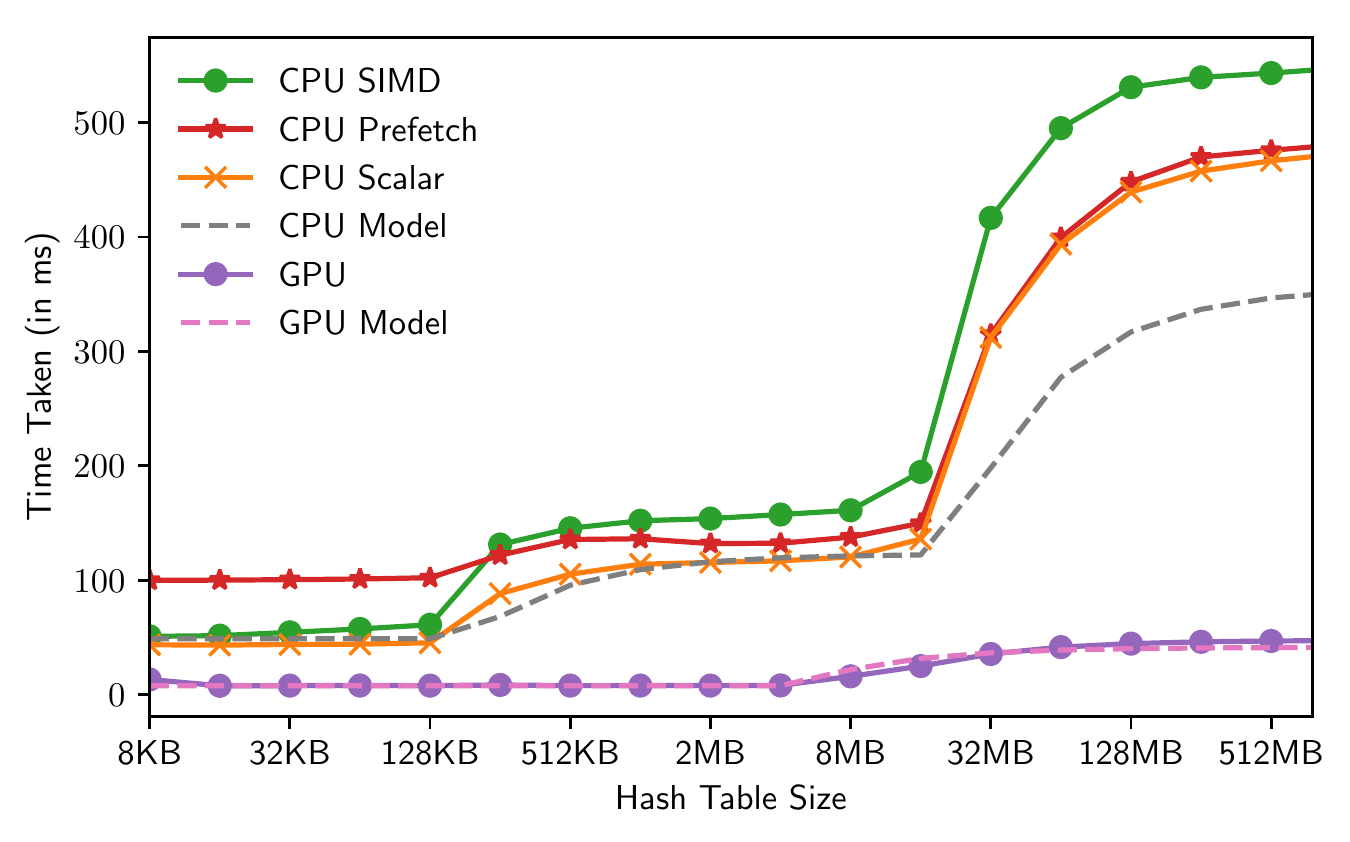}
  \vspace{-.3in}
  \caption{Join Performance.} 
  \label{fig:join-perf}
  \vspace{-.2in}
\end{figure}

\vspace{.05in}
\noindent\textbf{Performance Evaluation:} Figure~\ref{fig:join-perf} shows the performance evaluation of different implementations of Join. Both CPU and GPU variants exhibit step increase in runtime when the hash table size exceeds the cache size of a particular level. On the CPU, the step increases happen when the hash table size exceeds 256KB (L2 cache size) and 20MB (L3 cache size). On the GPU, the step increase happens when the hash table size exceeds 6MB (L2 cache size).

We see that \texttt{CPU SIMD} performs worse than \texttt{CPU Scalar}, even when the hash table is cache-resident. \texttt{CPU-SIMD} uses AVX2 instructions with 256-bit registers which represent 8 lanes of 32-bit integers. With 8 lanes, we process 8 keys at a time. However, a single SIMD gather used to fetch matching entries from the hash table can only fetch 4 entries at a time (as each hash table lookup returns an 8 byte slot. i.e., 4-byte key and 4-byte value, with 4 lookups filling the entire register). As a result, for each set of 8 keys, we do 2 SIMD gathers and then de-interleave the columns into to 8 keys and 8 values. 
This added overhead of extra instructions does not exist in the scalar version. \texttt{CPU SIMD} is also brittle and not easy to extend to cases where hash table slot size is larger than 8 bytes. Note that past work has evaluated vertical vectorization with key-only build relations which do not exhibit this issue~\cite{polychroniou2015rethinking,menon2017relaxed}. Comparing \texttt{CPU Prefetch} to \texttt{CPU Scalar} shows that there is limited improvement from prefetching when data size is larger than the L3 cache size. When the hash table fits in cache, prefetching degrades the performs due to added overhead of the prefetching instructions.

Due the step change nature of the performance curves, the ratio of the runtimes varies based on hash table size. When the hash table size is between 32KB and 128KB, the hash table fits in L2 on both CPU and GPU. In this segment, we observe that the runtime is bound by DRAM memory bandwidth on CPU and L2 cache bandwidth on the GPU. The average gains are roughly $5.5\times$ which is in line with the model.  When the hash table size is between 1MB and 4MB, the hash table fits in the L2 on the GPU and in the L3 cache on the CPU. The ratio of runtimes in this segment is $14.5\times$ which is the ratio of L2 cache bandwidth on GPU to the L3 cache bandwidth on the CPU. Finally when the hash table size is larger than 128MB, the hash table does not fit in cache on either GPU or CPU. The granularity of reads from global memory is 128B on GPU while on CPU it is 64B. Hence, random accesses into the hash table read twice the data on GPU compared to CPU. Given the bandwidth ratio is $16.2$x, we would expect it as roughly $8.1$x, however it is $10.5$x due to memory stalls. 
The fact that actual CPU results are slower than \texttt{CPU Model} is because the model assumes maximum main memory bandwidth, which is not achievable as the hash table causes random memory access patterns. 

\noindent \textbf{Discussion:}
The runtime of the build phase in the microbenchmark shows a linear increase with size of the build relation. The build phase runtimes are less affected by caches as writes to hash table end up going to memory. 

In this section, we modeled and evaluated the no partitioning join. Another variant of hash join is the partitioned hash join. Partitioned hash joins use a partitioning routine like radix partitioning to partition the input relations into cache-sized chunks and in the second step run the join on the corresponding partitions. Efficient radix-based hash join algorithms (\textit{radix join}) have been proposed for CPUs~\cite{balkesen2013main,blanas2011design,balkesen2013multi,chen2007improving} and for the GPUs~\cite{rui2017fast,sioulas2019hardware}. Radix join requires the entire input to be available before the join starts and as a result intermediate join results cannot be pipelined. Hence, while radix join is faster for a single join, radix joins are not used for queries with multiple joins. \new{While we do not explicitly model/evaluate radix joins, in the next section we discuss the radix partitioning routine that is the key component of such joins. That discussion shows that a careful radix partition implementation on both GPU and CPU are memory bandwidth bound, and hence the performance difference is roughly equal to the bandwidth ratio.}




\subsection{Sort}

\rev{In this section, we evaluate the performance of sorting 32-bit key and 32-bit value arrays based on the key. 
According to literature, the fastest sort algorithm for this workload is the radix sort algorithm. We start by describing the Least-Significant-Bit (LSB) radix sort on the CPU~\cite{polychroniou2014comprehensive} and on the GPU~\cite{merrill2011high}. LSB radix sort is the fastest for the workload on the CPU. We describe why the LSB radix sort does poorly in comparison on the GPU and why an alternate version called Most-Significant-Bit (MSB) radix sort does better on the GPU~\cite{gpu-sort2}. We present a model for the runtime of the radix sort algorithm and then analyze the performance characteristics of radix partitioning on CPU vs GPU. Our implementations are primarily based on previous work but this is first time that these algorithm are compared to each other.}


\rev{The LSB radix sort internally comprises a sequence of radix partition passes. Given an array $A$, radix $r$, and start bit $a$, a radix partition pass partitions the elements of the input array $A$ into a contiguous array of $2^r$ output partitions based on value of r-bits $e[a:a+r)$ (i.e., radix) of the key $e$. Both on the CPU and GPU, radix partitioning involves two phases. In the first phase (\textit{histogram phase}), each thread (CPU) / thread block (GPU) computes a histogram over its entries to find the number of entries in each partition of the $2^r$ partitions. In the second phase (\textit{data shuffling phase}), each thread (CPU) / thread block (GPU) maintains an array of pointers initialized by the prefix sum over the histogram and writes entries to the right partition based on these offsets. The entire sorting algorithm contains multiple radix partition passes, with each pass looking at a disjoint sets of bits of the key $e$ starting from the lowest bits $e[0:r)$ to highest bits $e[k-r:k)$ (where $k$ is the bit-length of the key).}

\rev{On the CPU, we use the implementation of Polychroniou et al.~\cite{polychroniou2014comprehensive}. In the histogram phase, each thread makes one pass over its partition, for each entry calculating its radix value and incrementing the count in the histogram (stored in the L1 cache). 
For the shuffle phase, we first compute a prefix sum over the histograms of all the threads (a 2D array of dimension $2^r \times t$ where t is the number of threads) to find the partition offsets for each of the threads. Next, each thread makes a pass over its partition using gathers and scatters to increment the counts in its offset array and writing to right offsets in output array. The implementation makes a number of optimizations to achieve good performance. 
Interested reader can refer to ~\cite{polychroniou2014comprehensive} for more details.}


\rev{On the GPU, we implemented LSB radix sort based on the work of Merrill et al.~\cite{merrill2011high}. In the histogram phase, each thread block loads a tile, computes a histogram that counts the number of entries in each partition, and writes it out to global memory. Prefix sum is used to find the partition offsets for each thread block in the output array. Next, in the shuffling phase each thread block reads in its offset array. The radix partitioning pass described above need to do stable partitioning i.e., ensures that for two entries with the same radix, the one occurring earlier in the input array also occurs earlier in the output array. Now on the GPU, in order to ensure stable partitioning for LSB radix sort we need to internally generate an offsets array for each thread from the the thread block offset array. For an r-bit radix partitioning, we need $2^r$ size histogram per thread. A number of optimizations have been proposed to store the histogram efficiently in registers, details of which are described in~\cite{merrill2011high}. Due to restriction on number of registers available per thread, stable radix partitioning pass can only process 7-bits at a time.}

\rev{Recently, Stehle et al.~\cite{gpu-sort2} presented an MSB radix sorting algorithm for the GPU. The MSB radix sort does not require stable partitioning. As a result, in the shuffle phase, we can just maintain a single offset array of size $2^r$ for the entire thread block. This allows MSB radix sort to process up to 8-bits at a time. Hence, the MSB radix sort to sort array of 32-bit keys with 4 passes each processing 8-bits at a time. On the other hand, LSB radix sort can processes only 7-bits at a time, and hence needs 5 radix partitioning passes processing 6,6,6,7,7 bits each.}

\begin{figure}[t]
  \centering
  \begin{subfigure}[b]{\columnwidth}
  \centering
  \includegraphics[width=0.95\columnwidth]{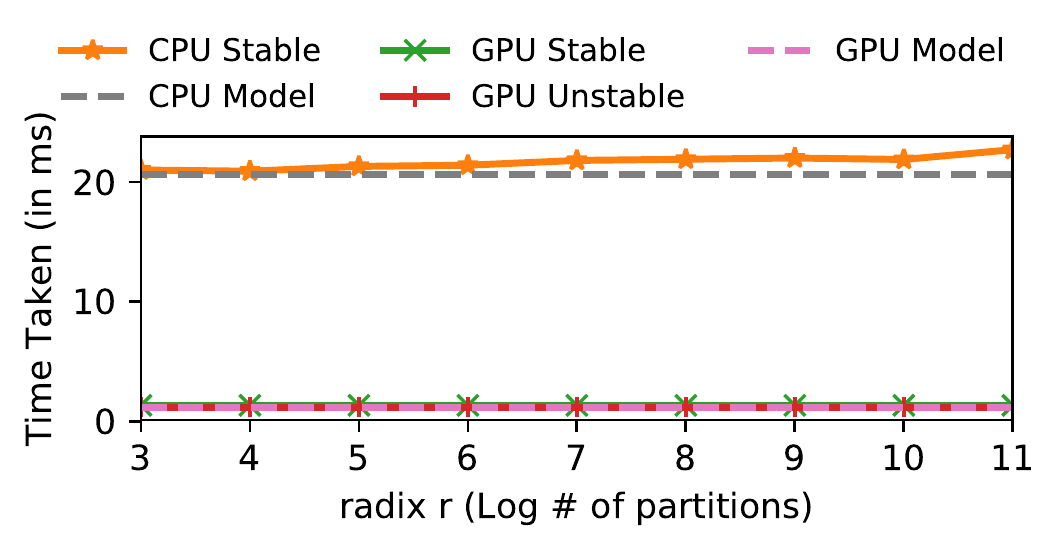}
  \vspace{-.1in}
  \caption{Radix histogram on CPU and GPU}
  \label{fig:radix-histogram}
  \end{subfigure}
  \vspace{-.1in}

  \begin{subfigure}[b]{\columnwidth}
  \centering
  \includegraphics[width=0.95\columnwidth]{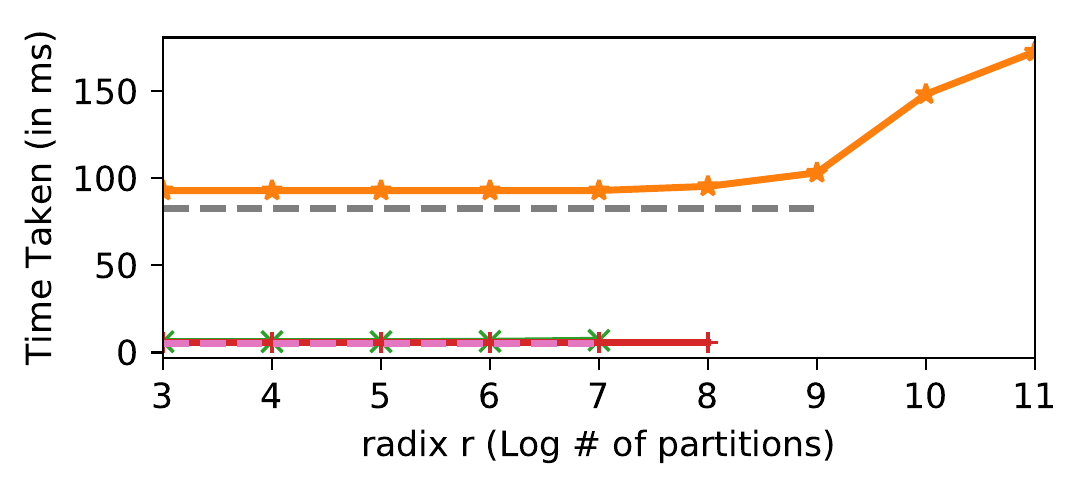}
  \vspace{-.1in}
  \caption{Radix shuffling on CPU and GPU}
  \label{fig:radix-shuffle}
  \vspace{-.1in}
  \end{subfigure}
  \caption{Sort Microbenchmark}
  \vspace{-.2in}
\end{figure}


\noindent \textbf{Model:} In the histogram phase, we read in the key column and write out a tiny histogram. The expected runtime is:
\vspace{-0.05in}
\begin{align*}
\textit{runtime}_{\textit{histogram}} = \frac{4 \times R}{B_r}
\end{align*}
where $R$ is the size of the input array and $B_r$ is the read bandwidth. In the shuffle phase, we read both the key and payload column and at the end write out  the radix partitioned key and payload columns. If the step is memory bandwidth bound, the runtime is expected to be:
\vspace{-0.05in}
\begin{align*}
\textit{runtime}_{\textit{shuffle}} = \frac{2 \times 4 \times R}{B_r} + \frac{2 \times 4 \times R}{B_w}
\end{align*}
where $B_w$ is the write bandwidth.

\noindent\textbf{Performance Evaluation:} We evaluate the performance of histogram and shuffle phase of the three variants: \texttt{CPU Stable} (stable partitioning on CPU), \texttt{GPU Stable} (stable partitioning on GPU), and \texttt{GPU Unstable} (unstable partitioning on GPU). 
We set the size of the input arrays at 256 million entries and vary the number of radix bits we partition on. Figure~\ref{fig:radix-histogram} shows the results for the histogram phase. Note that in the histogram phase there is no difference between \texttt{GPU Stable} and \texttt{GPU Unstable}. The histogram pass is memory bandwidth bound on both the CPU and GPU. Figure~\ref{fig:radix-shuffle} shows the results for the shuffle phase. \texttt{GPU Stable} is able to partition  up to 7-bits at a time whereas \texttt{GPU Unstable} is able to partition 8-bits at a time. \texttt{CPU Stable} is able to partition up to 8-bits a time while remaining bandwidth bound. Beyond 8-bits, the size of the partition buffers needed exceeds the size of L1 cache and the performance starts to deteriorate. 

Now that we have the radix partitioning passes, we look at the sort runtime. On the CPU, we use stable partitioning to implement LSB radix sort. It ends up running 4 radix partitioning passes each looking at 8-bits at time. On the GPU, MSB radix sort also sorts the data with 4 passes each processing 8-bits at a time. The time taken to sort $2^{28}$ entries is $464ms$ on the CPU and $27.08ms$ on the GPU. The runtime gain is $17.13 \times$ which is close to the bandwidth ratio of $16.2 \times$.

\begin{figure}
\centering
\begin{subfigure}[b]{0.45\columnwidth}
\begin{verbatim}
for each y in R:
  if y > v:
    output[i++] = v
\end{verbatim}
\caption{With branching}
\end{subfigure}
    ~ 
\begin{subfigure}[b]{0.45\columnwidth}
\begin{verbatim}
for each y in R:
  output[i] = y
  i += (y > v)
\end{verbatim}
\caption{With predication}
\end{subfigure}
\caption{Implementing selection scan}
\label{fig:selscan-example}
\vspace{-.1in}
\end{figure}


\section{Workload Evaluation}
\label{sec:evaluation}

\rev{Now that we have a good understanding of how individual operators behave on both CPU and GPU, we will evaluate the performance of a workload of full SQL queries on both hardware platforms. 
We first describe the query workload we use in our 
evaluation. We then present a high-level comparison of the performance of queries running on GPU implemented with the tile-based execution model versus our own equivalent implementation of the queries on the CPU. We also report the performance of Hyper~\cite{hyper} on CPU and Omnisci~\cite{omnisci} on the GPU which are both state-of-the-art implementations. As a case study, we provide a detailed performance breakdown of one of the queries to explain the performance gains. Finally, we present a dollar-cost comparison of running queries on CPU and GPU.}

We use two platforms for our evaluation. For experiments run on the CPU, we use a machine with a single socket Skylake-class Intel i7-6900 CPU with 8 cores that supports AVX2 256-bit SIMD instructions. For experiments
run on the GPU, we use an instance which contains an Nvidia V100 GPU. 
We measured the bidirectional PCIe transfer bandwidth to be 12.8GBps. More details of the two instances
are shown in Table 2. Each system is running on Ubuntu 16.04 and the GPU instance has CUDA 10.0. In our evaluation, we
ensure that data is already loaded into the respective device's memory before experiments start. 
We run each experiment 3 times and report the average measured execution time.



\vspace{-0.1in}
\subsection{Workload}
\label{sec:ssb-workload}
For the full query evaluation, we use the Star Schema Benchmark (SSB)~\cite{ssb} which has been widely used in various 
data analytics research studies~\cite{hippogriffdb,yinyang,gpudb2,funke2018pipelined}. SSB is a simplified version of the more popular TPC-H benchmark. It has 
one fact table \textit{lineorder} and four dimension tables \textit{date, supplier, customer, part} which are organized
in a star schema fashion. 
There are a total of 13 queries in the benchmark, divided into 4 query 
flights. In our experiments we run the benchmark with a scale factor of 20 which will generate the fact table with 120 million
tuples. The total dataset size is around 13GB.

\vspace{-0.1in}
\subsection{Performance Comparison}
\label{sec:perf-comparison}

In this section, we compare the query runtimes of benchmark queries implemented using block-wide functions on the GPU (\texttt{Standalone GPU}) to 
an equivalent efficient implementation of the query on the CPU (\texttt{Standalone CPU}). We also compare against Hyper (\texttt{Hyper}), a state-of-the-art OLAP DBMS and Omnisci (\texttt{Omnisci}), a commercial GPU-based OLAP DBMS.

In order to ensure a fair comparison across systems, we dictionary encode the string columns into integers prior to data loading
and manually rewrite the queries to directly reference the dictionary-encoded value. For example, a query with predicate
\texttt{s\_region = `ASIA'} is rewritten with predicate \texttt{s\_region = 2} where \texttt{2} is the dictionary-encoded value of \texttt{`ASIA'}. Some columns have a small number 
of distinct values and can be represented/encoded with 1-2 byte values. However, in our benchmark we make sure all column entries are 4-byte values 
to ensure ease of comparison with other systems and avoid implementation artifacts. Our goal is to understand the nature of the performance gains of equivalent implementations on GPU and CPU, 
and not to achieve best storage layout. We store the data in columnar format with each column represented as an array of 4-byte values.
On the GPU, we use a thread block size of 256 with tile size of 2056 (= 8$\times$256) resulting in 8 entries per thread per tile.

\begin{figure*}[t!]
  \centering
  \includegraphics[width=\linewidth]{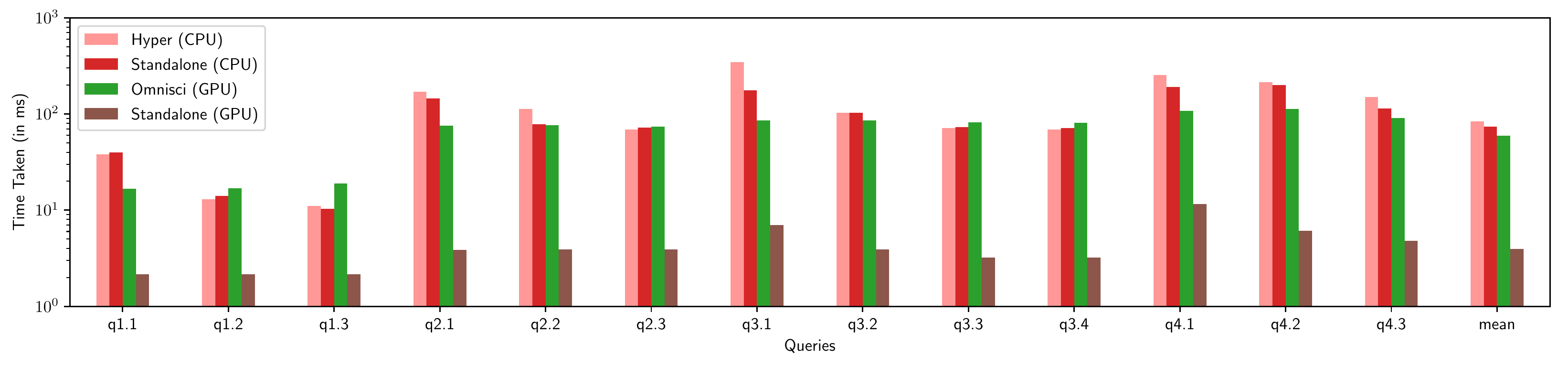}
  \vspace{-.3in}
  \caption{Star Schema Benchmark Queries}
  \label{fig:all-queries}
  \vspace{-.1in}
\end{figure*}

Figure~\ref{fig:all-queries} shows the results. Comparing \texttt{Standalone CPU} to \texttt{Hyper} shows that the former does on an average $1.17$x better than the latter. We believe \texttt{Hyper} is missing vectorization opportunities and using a different implementation of hash tables. The comparison shows that our implementation is a fair comparison and it is quite competitive compared to a state-of-the-art OLAP DBMS. We also compared against MonetDB~\cite{monetdb}, a popular baseline for many of the past works on GPU-based databases. We found that the \texttt{Standalone CPU} is on an average $2.5\times$ faster than MonetDB. 
We did not include it in the figure as it made the graph hard to read. We also tried to compare against Pelaton with relaxed-operator fusion~\cite{menon2017relaxed}. We found that the system could not load the scale factor 20 dataset. Scaling down to scale factor 10, its queries were significantly slower ($>$5$\times$) than Hyper or our approach.

\begin{table}[t]
\centering
\footnotesize
\begin{tabular}{|l|l|l|}
\hline
Platform & CPU & GPU \\
\hline
Model & Intel i7-6900 & Nvidia V100 \\
\hline
Cores & 8 (16 with SMT) & 5000 \\
\hline
Memory Capacity & 64 GB & 32 GB \\
\hline
L1 Size & 32KB/Core & 16KB/SM \\
\hline
L2 Size & 256KB/Core & 6MB (Total)\\
\hline
L3 Size & 20MB (Total)& - \\
\hline
Read Bandwidth & 53GBps & 880GBps \\
\hline
Write Bandwidth & 55GBps & 880GBps \\
\hline
L1 Bandwidth & - & 10.7TBps\\
\hline
L2 Bandwidth & - & 2.2TBps\\
\hline
L3 Bandwidth & 157GBps & -\\
\hline
\end{tabular}
\caption{Hardware Specifications}
\vspace{-.4in}
\label{tab:specs}
\end{table}

\rev{Comparing \texttt{Standalone GPU} to \texttt{Omnisci}, we see that our GPU implementation does significantly better than \texttt{Omnisci} with an average improvement of around $16\times$. Both methods run with the entire working set stored on the GPU. \texttt{Omnisci} treats each GPU thread as an independent unit. As a result, it does not realize benefits of blocked loading and better GPU utilization got from using the tile-based model. The comparison of \texttt{Standalone GPU} against \texttt{Omnisci} and \texttt{Standalone CPU} to \texttt{Hyper} serve as a sanity check and show that our query implementations are quite competitive.}

Comparing \texttt{Standalone GPU} to \texttt{Standalone CPU}, we see that the \texttt{Standalone GPU} is on average $25\times$ faster than the CPU implementation. This is higher than the bandwidth ratio of $16.2$. This is surprising given that in Section~\ref{sec:opcomp} we saw that individual query operators had a performance gain equal to or lower than the bandwidth ratio. \rev{The key reason for the performance gain being higher than the bandwidth ratio is the better latency hiding capability of GPUs. To get a better sense for the runtime difference, in the next subsection we discuss models for the full SQL queries and dive into why architecture differences leads to significant difference in performance gain from the bandwidth ratio.}

\vspace{-0.1in}
\rev{\subsection{Case Study}}

The queries in the Star Schema Benchmark can be broken into two sets: 1) the query flight $q1.x$ consists of queries with selections directly on the fact table with no joins and 2) the query flights $q2.x$, $q3.x$, $q4.x$ consist of queries with no selections on fact table and multiple joins --- some of which are selective. In this section, we analyze the behavior q2.1 in detail as a case study. 
Specifically, we build a model assuming the query is memory-bandwidth bound, derive the expected runtime based on the model, compare them against the observed runtime, and explain the differences observed.

Figure~\ref{fig:q21-query} shows the query: it joins the fact table \texttt{lineorder} with 3 dimension tables: \texttt{supplier}, \texttt{part}, and \texttt{date}. The selectivity of predicates on \texttt{p\_category} and \texttt{s\_region} are $1/25$ and $1/5$ respectively. The subsequent join of \texttt{part} and \texttt{supplier} have the same selectivity. We choose a query plan where \texttt{lineorder} first joins \texttt{supplier}, then \texttt{part}, and finally \texttt{date}, this plan delivers the highest performance among the several promising plans that we have evaluated. 

\begin{figure}[t]
\footnotesize
\begin{verbatim}
SELECT SUM(lo_revenue) AS revenue, d_year, p_brand
FROM lineorder, date, part, supplier
WHERE lo_orderdate = d_datekey
AND lo_partkey = p_partkey AND lo_suppkey = s_suppkey
AND p_category = 'MFGR#12' AND s_region = 'AMERICA'
GROUP BY d_year, p_brand
\end{verbatim}
\vspace{-.1in}
\caption{SSB Query 2.1}
\vspace{-.2in}
\label{fig:q21-query}
\end{figure}


The cardinalities of the tables \texttt{lineorder}, \texttt{supplier}, \texttt{part}, and \texttt{date} are $120M$, $40k$, $1M$, and $2.5k$ respectively. The query runs build phase for each of the 3 joins to build their respective hash tables. Then a final probe phase runs the joins pipelined. Given the small size of the dimension tables, the build time is much smaller than the probe time, hence we focus on modeling the probe time. On the GPU, each thread block processes a partition of the fact table, doing each of the 3 joins sequentially and updating a global hash table at the end that maintains the aggregate. 
Past work~\cite{mei2016dissecting} has shown that L2 cache on the GPU is an LRU set associative cache. Since hash tables associated with the \texttt{supplier} and \texttt{date} table are small, we can assume that they remain in the L2 cache. The size of the \texttt{part} hash table is larger than L2 cache. 
We model the runtime as consisting of 3 components:

\noindent 1) The time taken to access the columns of the fact table:
\begin{align*}
r_1 &= (\frac{4|L|}{C} + min(\frac{4|L|}{C}, |L|\sigma_1) + min(\frac{4|L|}{C}, |L|\sigma_1\sigma_2) \\
  & + min(\frac{4|L|}{C}, |L|\sigma_1\sigma_2)) \times \frac{C}{B_r}
\end{align*}
where $\sigma_1$ and $\sigma_2$ are join selectivities associated with join with \texttt{supplier} and \texttt{part} tables respectively, $|L|$ is the cardinality of the \texttt{lineorder} table, $C$ is size of cache line, and $B_r$ is the global memory read bandwidth. For each column except the first, the number of cache lines accessed is the minimum of: 1) accessing all cache lines of the column ($\frac{4|L|}{C}$) and 2) accessing a cache line per entry read ($|L|\sigma$). 

\noindent 2) Time taken to probe the join hash tables:
\begin{align*}
r_2 &= (2 \times |S| + 2 \times |D| + (1 - \pi) (|L|\sigma_1)) \times \frac{C}{B_r}
\end{align*}
where $|S|$ and $|D|$ are cardinalities of the \texttt{supplier} and \texttt{date} table, $(|L|\sigma_1)$ represents the number of lookups into the \texttt{part}hash table and $\pi$ is the probability of finding the \texttt{part} hash table lookup in the L2 cache.

\noindent 3) Time taken to read and write to the result table:
\begin{align*}
r_3 &= |L|\sigma_1\sigma_2 \times \frac{C}{B_r} + |L|\sigma_1\sigma_2 \times \frac{C}{B_w}
\end{align*}

The total runtime on GPU is $r_1 + r_2 + r_3$. The key difference with respect to CPU is that on the CPU, all three hash tables fit in the L3 cache. Hence for CPU, we would have $r_2 = (2 \times |S| + 2 \times |D| + 2 \times |P|)$. To calculate $\pi$, we observe that the size of the \texttt{part} hash table (with perfect hashing) is $2 \times 4 \times 1M = 8MB$. With the \texttt{supplier} and \texttt{date} table in cache, the available cache space is $5.7$MB. Hence the probability of \texttt{part} lookup in L2 cache is $\pi = 5.7/8$. Plugging in the values we get the expected runtimes on the CPU and GPU as 47~ms and 3.7~ms respectively compared to actual runtime of 125~ms and 3.86~ms. 


We see that the model predicted runtime on the GPU is close to the actual runtime whereas on the CPU, the actual runtime is higher than the modeled runtime. This is in large part because of the ability of GPUs to hide memory latency even with irregular accesses. SIMT GPUs run scalar code, but they ``tie'' all the threads in a \textit{warp} to execute the same instruction in a cycle. For instance, gathers and scatter are written as scalar loads and stores to non-contiguous locations. In a way, CPU threads are similar to GPU warps and GPU threads are similar to SIMD lanes. A key difference between SIMT model on GPU vs SIMD model on CPU is what happens on memory access. On the CPU, if a thread makes a memory access, the thread waits for the memory fetch to return. If the cache line being fetched is not in cache, it leads to a memory stall. CPU have prefetchers to remedy this, but prefetchers do not work well with irregular access patterns like join probes. On the GPU, a single streaming multiprocessor (SM) usually has 64 cores that can execute 2 warps (64 threads) at any point. However, the SM can keep $>2$ warps active at a time. On Nvidia V100, each SM can hold $64$ warps in total with $2$ executing at any point in time. Any time a warp makes a memory request, the warp is swapped out from execution into the active pool and another warp that
is ready to execute ends up executing. Once the memory fetch returns, the earlier warp can resume executing at the next available executor cores. This is similar to swapping of threads on disk access on CPU. This key feature allows GPUs to avoid the memory stalls associated with irregular accesses as long as enough other threads are ready to execute. Modeling query performance of multi-join queries on CPUs is an interesting open problem which we plan to address as future work.

\vspace{-0.1in}
\subsection{Cost Comparison}

The paper has so far demonstrated that GPUs can have superior performance than CPUs for data analytics.
However, GPUs are known to be more expensive than CPUs in terms of cost. 
Table~\ref{tab:cost} shows both the purchase and renting cost of CPU and GPU that match the hardware used in this paper (i.e., Table~\ref{tab:specs}). For renting costs, we use the cost of EC2 instances provided by Amazon Web Services (AWS). For CPU, we choose the instance type \texttt{r5.2xlarge} which contains a modern Skylake CPU with 8 cores, with a cost of $\$0.504$ per hour. For GPU, we choose the instance type \texttt{p3.2xlarge} whose specs are similar to \texttt{r5.2xlarge} plus it has an Nvidia V100 GPU, with a cost of $\$3.06$ per hour. The cost ratio of the two systems is about $6\times$. For purchase costs, we compare the estimate of a single socket server blade to the same server blade with one Nvidia V100 GPU. The cost ratio of the two systems at the high end is less than $6\times$. The average performance gap, however, is about $25\times$ according to our evaluation (cf. Section~\ref{sec:perf-comparison}), which leads to a factor of $4$ improvement in cost effectiveness of GPU over CPU. 
Although the performance and cost will vary a lot across different CPU and GPU technologies, the ratio between the two will not change as much. 
Therefore, we believe the analysis above should largely apply to other hardware selection. 

\begin{table}[t!]
  \centering
  \begin{tabular}{l|l|l|}
        & Purchase Cost & Renting Cost \\ \hline
    CPU & \$2-5K        & \$0.504 per hour \\ 
    GPU & \$CPU + 8.5K  & \$3.06 per hour 
  \end{tabular}
  \caption{Purchase and renting cost of CPU and GPU.}
  \label{tab:cost}
  \vspace{-.3in}
\end{table}


\vspace{-0.1in}
\subsection{Discussion}

\new{
In this paper, we showed through our model-based analysis and empirical evaluation that
there is limited gain from using GPUs as a coprocessor and that the runtime gain from running queries on the GPU vs  CPU is $1.5x$ the bandwidth ratio
of the two devices. We believe that these results should help pivot the community towards treating GPUs as primary execution engine.
However, this paper largely focused on using a single GPU, which has limited memory capacity. There are many challenges that need to be addressed
before GPUs have widespread adoption that were beyond the scope of this paper and make for exciting future work:
}

\begin{itemize} [leftmargin=*]
\item \new{\textbf{Distributed+Hybrid} It is possible to attach multiple GPUs onto a single machine that can greatly increase the aggregated HBM memory capacity. These machines will also having significant CPU memory. Executing queries on this heterogeneous system is still an open problem.}

\item \new{\textbf{Compression} Data compression could be used to fit more data into GPU's memory. GPUs have higher compute to bandwidth ratio than CPUs which could allow use of non-byte addressable packing schemes.}

\item \new{\textbf{Strings/Non-Scalar Data Types} Handling arbitrary strings and array data types efficiently on GPUs is still an open problem.}
\end{itemize}

\section{Conclusion}

This paper compared CPUs and GPUs on database analytics workloads. 
We demonstrated that running an entire SQL query on a GPU delivers better performance than using the GPU as an accelerator. 
To ease implementation of high-performance SQL queries on GPUs, we developed Crystal, a library supporting a tile-based execution model. 
Our analysis on SSB, a popular analytics benchmark, shows that 
modern GPUs are ~$25\times$ faster and ~$4\times$ more cost effective than CPUs. 
This makes a strong case for using GPUs as the primary execution engine when the dataset fits into GPU memory.



\newpage

{
\bibliographystyle{abbrv}

}

\end{document}